\documentclass{article}

\usepackage{PRIMEarxiv}

\usepackage[utf8]{inputenc} % allow utf-8 input
\usepackage{soul}

\usepackage{soul}
\usepackage{color}

\usepackage{booktabs} % in preamble
\usepackage{tabularx}

\usepackage{makecell} % for line breaks in cells
\usepackage[T1]{fontenc}    % use 8-bit T1 fonts
\usepackage{hyperref}       % hyperlinks
\usepackage{url}            % simple URL typesetting
\usepackage{booktabs}       % professional-quality tables
\usepackage{amsfonts}       % blackboard math symbols
\usepackage{nicefrac}       % compact symbols for 1/2, etc.
\usepackage{microtype}      % microtypography
\usepackage{lipsum}
\usepackage{fancyhdr}       % header
\usepackage{graphicx}       % graphics
\graphicspath{{media/}}     % organize your images and other figures under media/ folder
\usepackage{hyperref}
\usepackage{authblk}
\graphicspath{ {./images/} }

\usepackage{multicol}
\usepackage{multirow}
\usepackage{cite}
\usepackage{amsmath,amssymb,amsfonts}
\usepackage{algorithmic}
\usepackage{graphicx}
\usepackage{textcomp}
\usepackage{xcolor}
\usepackage{enumitem}
\usepackage{colortbl}
\usepackage{tikz}
\usepackage{hyperref}
\usepackage{geometry}
\usepackage{cleveref}
\usetikzlibrary{positioning}
\usepackage[edges]{forest}
\usepackage{xcolor}

\newcommand{\argmax}{\ensuremath{\mathop{\mathrm{argmax}}}}

\usepackage{xcolor} % simple, modern
\definecolor{mybrown}{RGB}{150,75,0}

\def\BibTeX{{\rm B\kern-.05em{\sc i\kern-.025em b}\kern-.08em
    T\kern-.1667em\lower.7ex\hbox{E}\kern-.125emX}}
\title{Non-Intrusive Automatic Speech Recognition Refinement: A Survey}

\author[1]{Mohammad Reza Peyghan}
\author[1]{Saman Soleimani Roudi}
\author[1]{Saeedreza Zouashkiani}
\author[1]{Sajjad Amini\thanks{Corresponding author: \quad Sajjad Amini, \texttt{s\_amini@sharif.edu}}
}
\author[2]{Fatemeh Rajabi}
\author[1]{Shahrokh Ghaemmaghami}
\affil[1]{Electronics Research Institute, Sharif University of Technology, Tehran, Iran}
\affil[2]{Department of Mathematics and Computer Science, Amirkabir University of Technology, Tehran, Iran}
\begin{document}
\maketitle

%% Abstract
\begin{abstract}
Automatic Speech Recognition (ASR) is an integral component of modern technology, powering applications such as voice-activated assistants, transcription services, and accessibility tools. Yet ASR systems continue to struggle with the inherent variability of human speech, such as accents, dialects, and speaking styles, as well as environmental interference, including background noise. Moreover, domain-specific conversations often employ specialized terminology, which can exacerbate transcription errors. These shortcomings not only degrade raw ASR accuracy but also propagate mistakes through subsequent natural language processing pipelines. Because redesigning an ASR model is costly and time-consuming, non-intrusive refinement techniques that leave the model's architecture intact have become increasingly popular. In this survey, we review current non-intrusive refinement approaches and group them into five classes: fusion, re-scoring, correction, distillation, and training adjustment. For each class, we outline the main methods, advantages, drawbacks, and ideal application scenarios. Beyond method classification, this work surveys adaptation techniques aimed at refining ASR in domain-specific contexts, reviews commonly used evaluation datasets along with their construction processes, and proposes a standardized set of metrics to facilitate fair comparisons. Finally, we identify open research gaps and suggest promising directions for future work. By providing this structured overview, we aim to equip researchers and practitioners with a clear foundation for developing more robust, accurate ASR refinement pipelines.
\end{abstract}

%% Keywords

% \end{frontmatter}


%%%%%%%%%%%%%%%%%%%%%%%%%%%%%%%%%%%%%%%%%%%%%%%%%%%%
%              I N T R O D U C T I O N
%%%%%%%%%%%%%%%%%%%%%%%%%%%%%%%%%%%%%%%%%%%%%%%%%%%%

\section{Introduction}

Automatic Speech Recognition (ASR) systems face significant challenges arising from variability in accents, dialects, and speaking styles, as well as environmental factors such as background noise. Moreover, domain-specific speech often includes specialized or rare terminology, which general-purpose ASR models tend to misrecognize, leading to transcription errors \cite{errattahi2018automatic}. These limitations reduce the overall performance and reliability of ASR systems, highlighting the need for techniques that can refine ASR models and enhance their output quality.
To address these challenges, researchers have explored four primary directions (D): (D1) designing comprehensive ASR architectures, (D2) training or fine-tuning existing models on new paired speech–text datasets, (D3) developing complementary models to enhance ASR outputs, and (D4) modifying training strategies to improve model performance.

\begin{figure*}[h]
    \centering
    \includegraphics[width=0.98\textwidth]{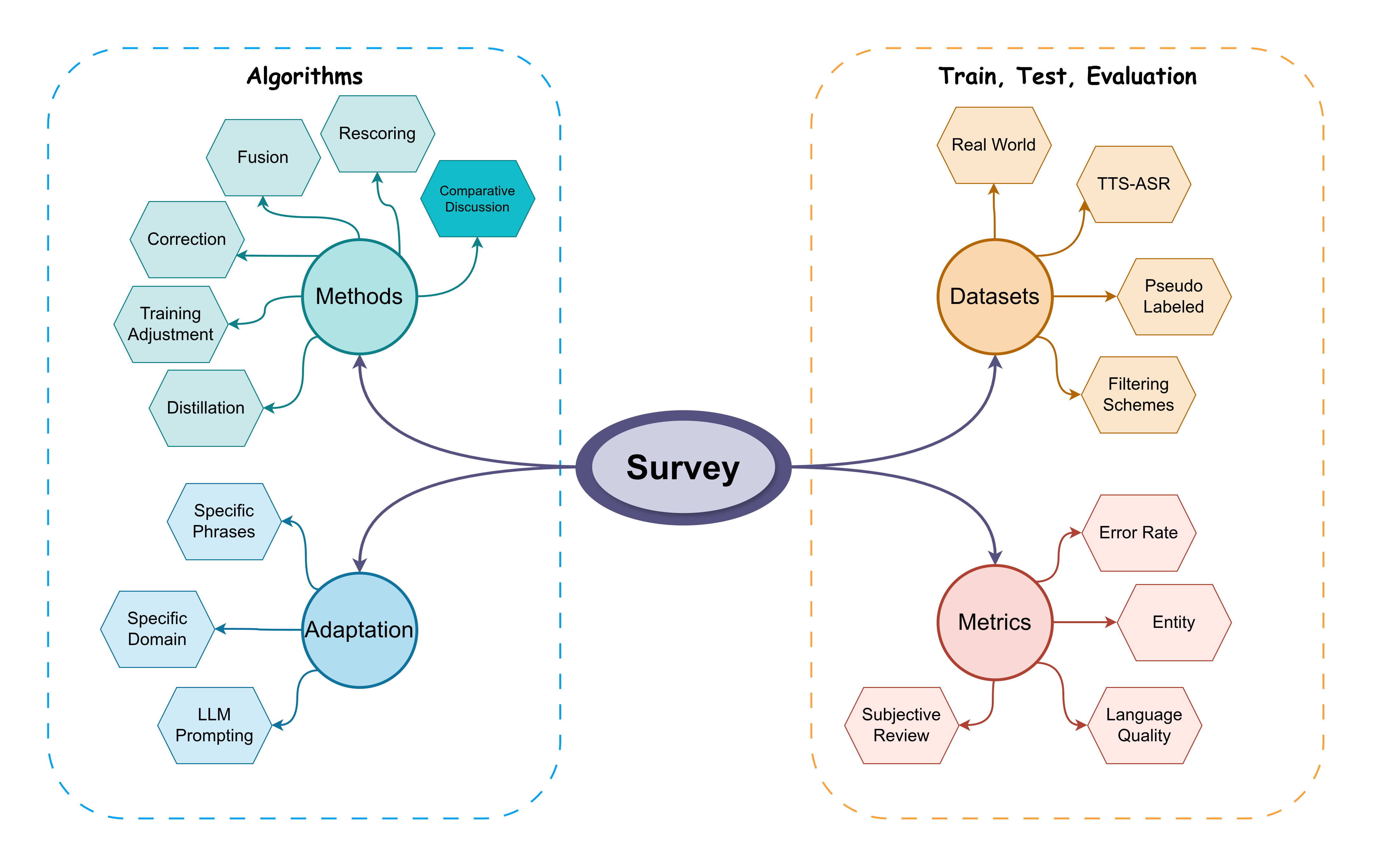}
    \caption{A comprehensive overview of survey sections and subsections.}
    \label{fig:image_1}
\end{figure*}

In D1 direction, with the advent of neural networks, particularly transformer-based architectures \cite{vaswani2017attention}, a diverse array of deep neural network (DNN) models has been proposed to improve the reliability and performance of ASR systems \cite{gulati2020conformer,baevski2020wav2vec,radford2023robust,meng2025dolphin}. Nevertheless, the design and development of novel architectures that deliver superior performance generally require substantial computational resources.

Regarding D2, the incorporation of large-scale paired speech–text datasets represents a valuable approach for enhancing ASR systems. However, such datasets remain limited, particularly for low-resource languages, posing significant challenges for ASR development. Furthermore, even when these datasets and their corresponding models are available, they are often inaccessible for modification, either because they are deployed as proprietary cloud-based services or because they require computational resources beyond what is typically available to end users. Additionally, training or fine-tuning existing models on new datasets entails a substantial risk of overfitting \cite{radford2023robust}. Taken together, these limitations frequently render such approaches impractical for real-world deployment.

For D3 and D4, researchers have developed methods to improve ASR performance without altering the neural architecture or requiring additional paired speech–text data. Conventional refinement approaches include rule-based post-processing \cite{Fiscus1997} and n-gram language models \cite{bassil2012asr}. However, rule-based methods do not generalize well, as it is infeasible to define an exhaustive set of rules capable of handling diverse linguistic variations. Similarly, n-gram approaches are constrained by their limited context window (typically up to 4–5 words), which restricts their ability to capture long-range dependencies, an essential aspect of natural language understanding (NLU) \cite{lenci2023understandingnaturallanguageunderstanding}.

With the advent of more powerful computing hardware and DNNs, which excel at learning generalizable representations, significant progress has been made in ASR refinement. Researchers began adopting Neural Language Models (NLMs), particularly Recurrent Neural Networks (RNNs) \cite{medsker1999recurrent}, to refine ASR outputs \cite{kannan2018}. Although RNNs can model longer contexts than n-grams, they often struggle to maintain information over extended sequences due to vanishing gradient and forgetting issues.

The introduction of transformers \cite{vaswani2017attention} marked a major milestone in ASR modeling and refinement, driving rapid innovation across both domains. Many studies have since incorporated transformer-based architectures into ASR pipelines in various forms \cite{huang2019empirical,zhang2019investigation,zhang2020spelling,hu2021transformer,shu2024error}, leveraging their ability to process all input tokens simultaneously and apply selective attention mechanisms. Nonetheless, early transformer-based methods \cite{xia2017deliberation,tanaka2021cross} suffered from high computational costs, particularly during decoding. To address this issue, several solutions were proposed, including encoder-only designs \cite{cheng2020spellgcn,zhang2020spelling,peyghan2025cmc}, parallel autoregressive (AR) decoding \cite{he2023ed}, and non-autoregressive (NAR) decoding strategies \cite{shu2024error,Wang2022Mar}.

More recently, the emergence of Large Language Models (LLMs) has profoundly influenced ASR refinement paradigms. A growing body of work explores integrating LLMs to enhance ASR output correction, with recent approaches employing Retrieval-Augmented Generation (RAG) to provide domain-relevant context and thereby improve correction accuracy \cite{pusateri2025retrieval,robatian2025gecrag}.

\definecolor{citeblue}{RGB}{219,238,254}
% Styles
\forestset{
  default box/.style={
    draw,
    rounded corners,
    align=left,
    inner sep=4pt,
    font=\small
  },
  cite box/.style={
    default box,
    fill=citeblue,
    font=\scriptsize
  }
}
\begin{figure*}[t]
\centering
\resizebox{\textwidth}{!}{%
\begin{forest}
  for tree={
    grow'=0, % left-to-right
    child anchor=west,
    parent anchor=east,
    anchor=west,
    edge path={
      \noexpand\path[\forestoption{edge}] (!u.parent anchor) -- +(5pt,0) |- (.child anchor)\forestoption{edge label};
    },
    l sep+=10pt,
    s sep+=2pt,
    align=left
  }
  [{Non-Intrusive ASR Refinement}, default box
    [Methods, default box
        [{Fusion}, default box
            [Shallow, default box
                [{~\cite{Gulcehre2015}, ~\cite{kannan2018}, ~\cite{McDermott2019}, ~\cite{Meng2021-estimation}, ~\cite{ma2023internal}, ~\cite{ogawa2023iterative}, ~\cite{Hori2025}}, cite box]
            ]
            [Deep, default box
                [{~\cite{Gulcehre2015}, ~\cite{mittal2024salsa}, ~\cite{zhang2025end}}, cite box]
            ]
            [Cold, default box
                [{~\cite{sriram2018cold}, ~\cite{Cho2019}, ~\cite{Kim2021}}, cite box]
            ]
        ]
        [{Rescoring}, default box
            [{\textbf{First-pass:} ~\cite{murthy2024initial}, ~\cite{deoras2011fast}, ~\cite{auli2013joint}, \cite{kumar2017lattice}, ~\cite{liu2016two}, ~\cite{sundermeyer2014lattice}, ~\cite{xu2018pruned} \\ \textbf{Second-pass:} \cite{huang2019empirical}, ~\cite{guo2019spelling}, \cite{gandhe2020audio}, ~\cite{ogawa2018rescoring}, \cite{sainath2019two}, ~\cite{li2020parallel}, \cite{hu2020deliberation}, \cite{hu2021transformer}, ~\cite{Hu2022Scaling}, ~\cite{shin2019effective}, ~\cite{salazar2020masked}, ~\cite{udagawa2022effect}, ~\cite{xu2022rescorebert}, ~\cite{chiu2021innovative}, \cite{ma2023can}, \cite{tur2024progres}, ~\cite{shivakumar2025speech} \\ \textbf{First-pass \& Second-pass:} \cite{tran1996word}}, cite box]
        ]
        [{Correction}, default box
            [Rule-Based, default box
                [{\cite{Fiscus1997}, \cite{Imai2025Saki}, \cite{bassil2012asr}, ~\cite{bassil2012post}, ~\cite{zhou2006multi}, \cite{DHaro2016Automatic}, \cite{Zietkiewicz2022}}, cite box]
            ]
            [NLM, default box
                [Encoder-Based, default box
                    [{\cite{hong2019faspell}, \cite{cheng2020spellgcn}, \cite{zhang2021correcting}, \cite{fan2023boosting}, \cite{zhang2020spelling}, \cite{chen2021integrated}, \cite{peyghan2025cmc}}, cite box]
                ]
                [Decoder-Inclusive, default box
                    [Autoregressive Decoders, default box
                        [{\cite{Tanaka2018NeuralErrorCorrective}, \cite{zhang2019investigation}, \cite{hrinchuk2020correction}, \cite{li2021boost}, \cite{zhao2021bart}, \cite{yeen2023learned}, \cite{ma2023nbest}, \\~\cite{dutta2022error}, ~\cite{hu2020deliberation}, ~\cite{tanaka2021cross}, ~\cite{li2024crossmodal}, ~\cite{jiang2024cross}, ~\cite{kumar2023visual}}, cite box]
                    ]
                    [Non-Autoregressive Decoders, default box
                        [{~\cite{leng2021fastcorrect}, ~\cite{guo2021global}, ~\cite{shen2022mask}, ~\cite{yang2022asr}, ~\cite{leng2021fastcorrect2}, ~\cite{leng2023softcorrect}, \\~\cite{fang2022non}, \cite{futami2022non}, \cite{zhang2023patcorrect}, ~\cite{wang2022effective},  \cite{du2022cross}, \cite{shu2024error}}, cite box]
                    ]
                ]
            ]
            [LLM, default box
                [{~\cite{min2023exploring}, ~\cite{yang2023generative}, \cite{ma2023can}, \cite{ma2025asr}, \cite{naderi2024towards}, ~\cite{pu2023multi}, \cite{sachdev2024evolutionary}, ~\cite{tang2025full}, \\ ~\cite{tang2025chain}, ~\cite{udagawa2024robust}, ~\cite{li2024pinyin}, ~\cite{li2024investigating}, ~\cite{Ghosh2024Jun}, \cite{radhakrishnan2023whispering}, \cite{chen2024s}, ~\cite{hu2024listen}}, cite box]
            ]
            [RAG-Integrated, default box
                [{~\cite{robatian2025gecrag}, ~\cite{ghosh2024failing}, ~\cite{li2024highprecision}, ~\cite{xiao2025contextual}}, cite box]
            ]
        ]
        [{Distillation}, default box
            [{~\cite{bai2019learn}, ~\cite{bai2019integrating}, ~\cite{futami2020distilling}, \cite{bai2021fast}, \cite{futami2022distilling}, \cite{deng2022improving}, \cite{hentschel2024keep}, \cite{choi2022distilll2s}, \cite{kubo2022knowledge}, ~\cite{lee2022knowledge}, \cite{han2023knowledge}}, cite box]
        ]
        [{Training Adjustment}, default box
            [ILMT, default box
                [{~\cite{meng2021internal}}, cite box]
            ]
            [MWE Training, default box
                [{~\cite{hori2016minimum}, \cite{prabhavalkar2018minimum}}, cite box]
            ]
            [Label Smoothing, default box
                [{~\cite{chorowski2016towards}}, cite box]
            ]
        ]
    ]
    [Adaptation, default box
        [Domain-Specific Training, default box
            [{\cite{mai2022unsupervised}, \cite{nanayakkara2022clinical}, ~\cite{Jia2025EPIC}}, cite box]
        ]
        [Capturing Specific Phrases or Words, default box
            [{\cite{sarma2004context}, \cite{anantaram2018repairing}, ~\cite{zhao2019shallow}, ~\cite{huang2025neural}, ~\cite{kolehmainen2023personalization}, \cite{wang2022towards}, ~\cite{he2023ed}, \\~\cite{he2025pmf}, \cite{antonova2023spellmapper}, ~\cite{pusateri2025retrieval}, ~\cite{wang2024dancer}, ~\cite{luo2025generative}, ~\cite{im2025deragec}, \cite{zhou2025autodrafting}}, cite box]
        ]
        [LLM-Based Domain Generalization, default box
            [{~\cite{ebadi2024extracting}, ~\cite{adedeji2024sound}, \cite{adedeji2025multicultural}}, cite box]
        ]
    ]
  ]
\end{forest}
}
\caption{Non-Intrusive Refinement Methods and Adaptation Techniques for Automatic Speech Recognition.}\label{fig:forest}
\end{figure*}

In parallel, several training-adjustment approaches have been proposed to enhance ASR reliability without modifying the model architecture or requiring additional paired data. Instead, these methods focus on optimizing training objectives and strategies to produce more robust systems \cite{futami2020distilling,hentschel2024keep,meng2021internal}.

%%%%%%%%%%%%%%%%%%%%%%

These techniques, which correspond to the D3 and D4 directions, either develop complementary models or modify ASR training strategies. They are referred to as \textbf{\textit{non-intrusive}} in this research because they improve ASR performance without altering its core architecture or requiring additional paired training data.
Instead, they leverage existing resources or external strategies, such as language models (LMs) or auxiliary training objectives, to improve system performance. In this survey, non-intrusive methods are categorized into five main classes:
\textit{Fusion} (combining outputs from ASR and LM models to improve accuracy),
\textit{Rescoring} (re-ranking hypotheses using LMs to select the most likely transcription),
\textit{Correction} (generating improved transcriptions by correcting errors in the initial output),
\textit{Distillation} (transferring knowledge from an external LM teacher to ASR model), and
\textit{Training adjustment} (optimizing the training process to improve the model's understanding of existing data).

These methods range from output-level correction techniques that operate on the final transcription and require minimal model access to approaches that distill external knowledge to retrain all or part of the ASR model’s parameters, requiring full model access.

While prior surveys have examined specific aspects of ASR refinement, including human-aided and conventional approaches \cite{errattahi2018automatic}, LM integration in encoder–decoder systems \cite{toshniwal2018comparison}, LLM-based Chinese error correction \cite{wei2024asr}, and AI-driven medical transcription enhancement \cite{saadat2025enhancing}, a comprehensive review of non-intrusive techniques is still lacking. Furthermore, existing studies often overlook the diversity of dataset construction schemes and evaluation protocols, which hinders consistent and standardized comparisons across methods.
This survey addresses these gaps through the following key contributions:
\begin{enumerate}
\item \textbf{Categorization of non-intrusive ASR refinement techniques}: A comprehensive taxonomy that classifies non-intrusive methods into five categories—fusion, rescoring, correction, distillation, and training adjustment.
\item \textbf{Analysis of domain-specific adaptation techniques}: An examination of methods designed for specific domains or terminology sets, distinguishing them from general-purpose non-intrusive approaches.
\item \textbf{Review of datasets and creation schemes}: An overview of datasets commonly used in ASR refinement, along with their construction processes and filtering strategies.
\item \textbf{Compilation of evaluation metrics}: A summary of standard evaluation metrics to support consistent comparison across methods.
\item \textbf{Identification of research gaps and future directions}: A discussion of open challenges, current limitations, and potential avenues for future research.
\end{enumerate}

% Structuring the paper
The remainder of this survey is organized as follows. Section~\ref{sec:methods} presents the non-intrusive methods, encompassing fusion, rescoring, correction, distillation, and training adjustment techniques. Section~\ref{sec:adaptation} discusses domain-specific adaptation methods that aim to improve ASR transcriptions within specialized domains or for restricted sets of phrases. Section~\ref{sec:datasets} reviews the datasets used in ASR refinement research, outlining their creation processes, filtering schemes, and data sources. Section~\ref{sec:metrics} introduces commonly used evaluation metrics in ASR refinement and highlights emerging metrics needed to enable more robust comparisons. Finally, Section~\ref{sec:research-gap} identifies existing research gaps and proposes future research directions. The overall structure of this study is illustrated in Figure\,\ref{fig:image_1}, and Figure\,\ref{fig:forest} provides a detailed presentation of non-intrusive techniques proposed for ASR refinement.

%%%%%%%%%%%%%%%%%%%%%%%%%%%%%%%%%%%%%%%%%%%%%%%%%%%%
%              M E T H O D S
%%%%%%%%%%%%%%%%%%%%%%%%%%%%%%%%%%%%%%%%%%%%%%%%%%%%

\begin{figure*}[h]
    \centering
    \includegraphics[width=0.9\textwidth, height=0.55\textwidth]{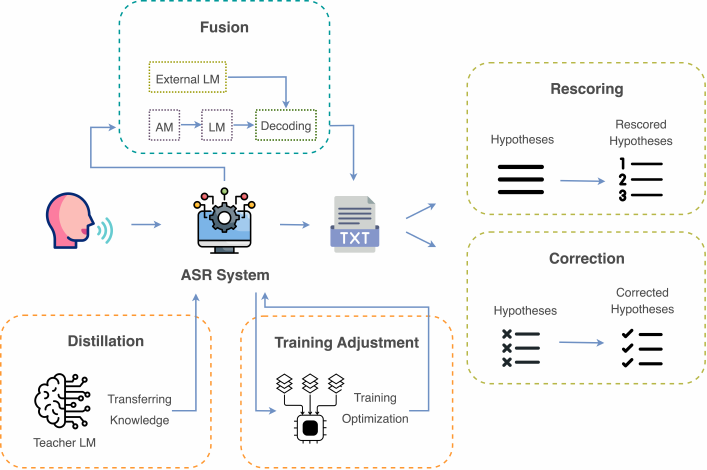}
    \caption{Schematic of ASR refinement methods (AM and LM refer to Acoustic Model and Language Model, respectively).}
    \label{fig:image_2}
\end{figure*}

\section{Methods}
\label{sec:methods}
In this section, we detail non-intrusive methods for refining ASR systems, classifying them into five main categories: 

\begin{itemize}
    \item \textbf{Fusion}: Integrating an external LM into the decoding step
    \item \textbf{ReScoring}: Re-evaluating the initial lattice or n-best hypotheses using an external scoring module (e.g., an LM)
    \item \textbf{Correction}: Generating a new transcript from the initial hypotheses
    \item \textbf{Distillation}: Retraining the ASR model with an external LM knowledge
    \item \textbf{Training adjustment}: Enhancing performance through novel training strategies without external knowledge
\end{itemize}

Figure \ref{fig:image_2} provides an overview of the main method categories discussed in this section. Additionally, Table \ref{tab:method-comparison} presents a side-by-side comparison of these methods, highlighting key practical considerations.

\subsection{Fusion}
\label{method-fusion}

This section reviews methods for \textit{fusing} external LMs into the decoding step of end‑to‑end (E2E) ASR systems. E2E architectures jointly learn acoustic, pronunciation, and language information but cannot exploit large unpaired text corpora without explicit fusion strategies \cite{prabhavalkar2023end}. 
A broad taxonomy of LM integration techniques is presented in \cite{toshniwal2018comparison}, comparing shallow fusion, deep fusion, cold fusion, LM-as-decoder-layer, and LM integration via multi-task training. The authors report that shallow fusion achieves the lowest first-pass word error rate (WER) (a detailed description of WER and other commonplace metrics in ASR systems is provided in Section \ref{sec:metrics}).
However, in this section we focus exclusively on shallow, deep, and cold fusion methods, reviewing non-intrusive approaches that, as mentioned earlier, neither alter the original ASR architecture nor require additional data. Moreover, multi-task training of the ASR system that follows the aforementioned conditions will be discussed in Section \ref{ta-ilmt}.
In this part, we first review shallow fusion techniques, then discuss deep fusion methods, and finally conclude with cold fusion approaches.

\subsubsection{Shallow Fusion}\label{method-shallow-fusion} 
\begin{figure*}[h]
    
    \centering
    \includegraphics[width=0.7\textwidth, height=0.28\textwidth]{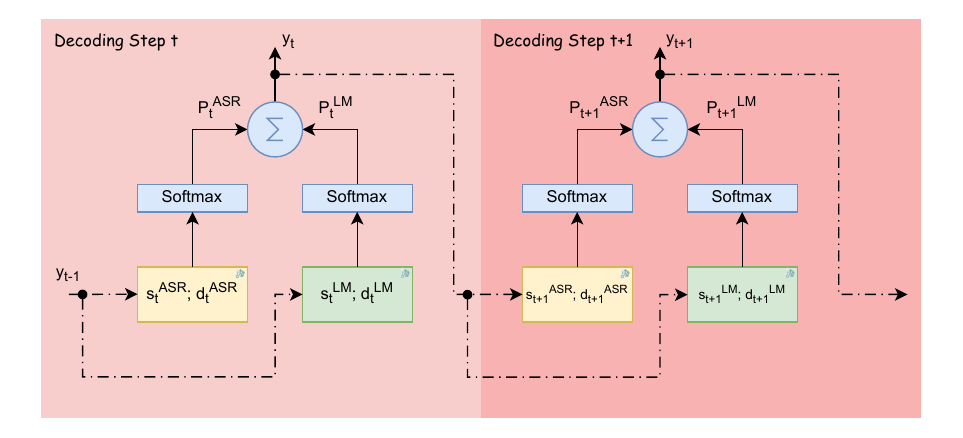}
    \caption{Schematic of Shallow Fusion in Two Consecutive Decoding Steps.}
    \label{fig:shallow}
\end{figure*}

Shallow fusion is an inference-only technique for end-to-end ASR systems in which an external language model (LM), trained separately on large text corpora, is integrated during beam-search decoding. This integration is typically performed via log-linear interpolation of the encoder--decoder model score (ASR score) $p_{\mathrm{ASR}}$ and the LM score $p_{\mathrm{LM}}$ at each time step~\(t\)~\cite{Gulcehre2015}, as illustrated in Figure \ref{fig:shallow}. Formally, given an acoustic input \(x\), hypotheses \(y\) are chosen to maximize

\begin{equation}
y^* = \argmax_{y_{1:T}}
\sum_{t=1}^{T}\Big[
\log p_{\mathrm{ASR}}(y_t \mid y_{<t}, x)
+ \lambda\,\log p_{\mathrm{LM}}(y_t \mid y_{<t})
+ \gamma \,\mathrm{penalty}(x,y_{1:t})
\Big]
\end{equation}

where \(y_{1:t}=(y_1,\dots,y_t)\), \(y_{<t}=(y_1,\dots,y_{t-1})\), and the hyperparameters \(\lambda\) and \(\gamma\) are tuned on the held-out data. The \(\mathrm{penalty}\) term is utilized to control hypothesis length by imposing a fixed cost per decoded token, thereby discouraging overly short or overly long transcriptions and balancing insertion/deletion errors. Several studies have refined shallow‑fusion decoding with targeted penalty terms \cite{Wu2016,chorowski2016towards,Gong2022}. 
% eval different archs and token sizes for accuracy and inference cost

Kannan et al.~\cite{kannan2018} systematically analyze shallow fusion for sequence-to-sequence (Seq2Seq) ASR by evaluating how different external LM architectures, various decoding units, and fusion‐weight settings affect inference‐time integration. Their study culminates in actionable guidelines for selecting LM types and decoding strategies to maximize shallow fusion gains in practical ASR deployments. They demonstrate the effectiveness of this simple fusion approach, which requires no ASR retraining and leverages large text-only corpora through LMs to improve first-pass decoding and ultimately reduce WER.

Shallow fusion effectively integrates external LMs during inference; yet, it is fundamentally a heuristic interpolation without a clear probabilistic foundation and remains biased toward the source-domain internal LM prior \cite{Meng2021-estimation,meng2021internal,McDermott2019,ma2023internal}. To reduce this bias, McDermott et al. \cite{McDermott2019} introduce the \emph{Density Ratio Method} (DRM), which applies Bayes’ rule to subtract the source‑domain score from the E2E posterior before adding the target‑domain LM score, thereby replacing the source prior with the desired external prior (an external LM is trained on the source domain). The DRM objective is:

\begin{equation}
y^* = \argmax_{y_{1:T}}
\sum_{t=1}^{T}\Big[
\log p(y_t \mid y_{<t}, x; \theta_{\mathrm{E2E}})
+ \lambda_{T}\,\log p_{\mathrm{LM}}^{\mathrm{target}}(y_t \mid y_{<t}; \theta^{\mathrm{target}}_{\mathrm{LM}})
- \lambda_{S}\,\log p_{\mathrm{LM}}^{\mathrm{source}}(y_t \mid y_{<t}; \theta^{\mathrm{source}}_{\mathrm{LM}})
\Big]
\end{equation}

where \(p_{\mathrm{LM}}^{\mathrm{target}}\) and \(p_{\mathrm{LM}}^{\mathrm{source}}\) denote the target and source LM probabilities, respectively. DRM consistently improves shallow fusion in cross‑domain evaluations.

Compared to the \textit{Density Ratio}, which subtracts a separately trained source-domain LM under a hybrid-style factorization, Meng et al.~\cite{Meng2021-estimation} proposed internal LM estimation (ILME). ILME employs the \textit{Joint Softmax Approximation} to estimate and subtract the E2E model's own internal LM (ILM) scores during inference. This approach achieves the following benefits: 
(i) eliminates the need for an external source LM, 
(ii) corrects the flawed Acoustic Model (AM)--LM decomposition (decoder of E2E models does not explicitly represent an LM in the source domain) in DRM objective function, and 
(iii) enables more accurate and efficient fusion. 
Equation \ref{eq:ILME} illustrates the decoding objective in ILME:

\begin{equation}
\label{eq:ILME}
y^* = \argmax_{y_{1:T}}
\sum_{t=1}^{T}\Big[
\log p(y_t \mid y_{<t}, x; \theta_{E2E})
+ \lambda_{T}\,\log p_{\mathrm{LM}}^{\mathrm{target}}(y_t \mid y_{<t}; \theta^{\mathrm{target}}_{LM})
- \lambda_{S}\,\log p_{\mathrm{LM}}^{\mathrm{source}}(y_t \mid y_{<t}; \theta_{E2E})
\Big]
\end{equation}

Moreover, Ma et al.~\cite{ma2023internal} observe that subtracting a well‑trained LM in standard DRM and ILME can degrade general‑domain ASR, especially when the ILM itself is top-performing; to mitigate this, they propose ILME based Adaptive Domain Adaptation (ILME-ADA), which at inference computes:

\begin{equation}
\label{eq:ADA_tok_argmax}
\begin{split}
y^* = \argmax_{y_{1:T}}\sum_{t=1}^{T}\Big[&
\log p_{\mathrm{ASR}}(y_t\mid y_{<t},x)
- \lambda_{\mathrm{ILM}} \log p_{\mathrm{ILM}}(y_t\mid y_{<t})\\
&\quad + \max\{\lambda \log p_{\mathrm{ELM}}(y_t\mid y_{<t}),\;
\lambda_{\mathrm{ILM}} \log p_{\mathrm{ILM}}(y_t\mid y_{<t})\}\Big]
\end{split}
\end{equation}

where ELM and ILM stand for external and internal language models, respectively. Hence, it dynamically selects the more reliable LM score at each decoding step \(t\) and achieves robust domain adaptation without hurting overall performance.

\begin{table*}[t]
\centering
\caption{Comparison of non-intrusive ASR refinement methods for real-world use at three levels: low, medium, and high.}\label{tab:method-comparison}
\resizebox{\textwidth}{!}{%
\begin{tabular}{lccccc}
\toprule
\thead{Aspect} & \thead{Fusion} & \thead{Rescoring} & \thead{Correction} & \thead{Distillation} & \thead{Training Adjustment} \\
\midrule
Trainability (ASR)
& \makecell[l]{shallow: Freeze \\ deep: Freeze \\ Cold: Trainable} 
& Freeze 
& Freeze 
& Trainable 
& Trainable \\
\midrule
Trainability (Refiner)
& \makecell[l]{shallow: Freeze \\ deep: Trainable \\ \small{(only gating params)} \\ Cold: Trainable \\ \small{(only gating params)}} 
& Freeze / Trainable 
& Trainable 
& Not Included 
& Not Included \\
\midrule
Implementation Complexity & Medium & Medium & Medium--High & High & Medium \\
\midrule
Pass Type & First-pass & \makecell[l]{lattice: first-pass \\ n-best: second-pass} & Second-pass & First-pass & First-pass \\
\midrule
Latency & High & Medium & Medium--High & Low (zero) & Low (zero) \\
\midrule
Inference Memory & Medium--High & Medium & Medium--High & Low (zero) & Low (zero) \\
\midrule
Robustness & High & Medium--High & Low--Medium & High & Medium \\
\bottomrule
\end{tabular}%
}
\end{table*}

Ogawa et al.~\cite{ogawa2023iterative} introduce \textit{Iterative Shallow Fusion} (ISF), in which an external \textit{backward language model} (BLM) is applied repeatedly to partial ASR hypotheses during inference, successively updating previous BLM scores at each decoding iteration. To tailor the BLM to partial sequences, they train a \textit{partial sentence--aware BLM} (PBLM) on reversed text that includes incomplete sentences, improving its efficacy within ISF. They further show that combining ISF with standard shallow fusion using a \textit{forward LM} (FLM) takes advantage of the complementary strengths of BLM and FLM, and that PBLM-based ISF alone achieves performance on par with FLM-based shallow fusion.

Recently, advances in LLMs have also motivated their integration within shallow fusion frameworks. Hori et al.~\cite{Hori2025} introduce delayed fusion, a shallow-fusion–style technique that delays LLM scoring until beam pruning at word boundaries is complete. This approach reduces the number of hypotheses evaluated and costly LLM inference calls, while enabling on-the-fly re-tokenization to handle ASR–LLM vocabulary mismatches. This deferred integration maintains inference latency on par with standard shallow fusion and surpasses the standard shallow fusion in error reduction.

\subsubsection{Deep Fusion}
\begin{figure}[h]
    
    \centering
    \includegraphics[width=0.7\textwidth, height=0.28\textwidth]{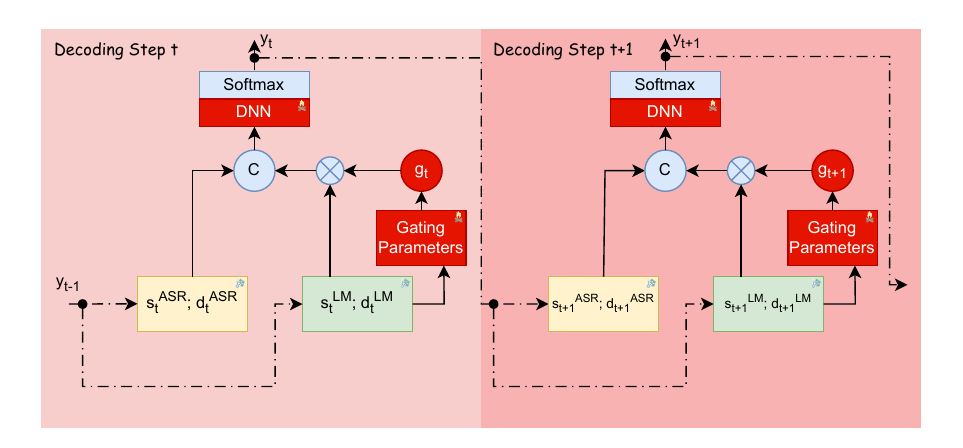}
    \caption{Schematic of Deep Fusion in Two Consecutive Decoding Steps.}
    \label{fig:deep}
\end{figure}

This method augments an ASR decoder with a pretrained external LM by learning a small gating network, while freezing both the ASR encoder–decoder and the LM weights during the training of the gating parameters \cite{Gulcehre2015}.
According to Figure \ref{fig:deep}, at each decoding step \(t\), a scalar gate \(g_t\) computes how much of the LM hidden state \(s_t^{\mathrm{LM}}\) to inject into the ASR decoder state \(s_t\):

\begin{equation}
  g_t = \sigma\bigl(v^\top s_t^{\mathrm{LM}} + b\bigr),
\end{equation}
where \(\sigma(\cdot)\) is the \textit{sigmoid} activation function and \((v\in\mathbb{R}^d,\,b\in\mathbb{R})\) are trainable \cite{Gulcehre2015,toshniwal2018comparison}. The fused decoder state is then

\begin{equation}
  s_t^{\mathrm{DF}} = \bigl[s_t;\;g_t\,\odot s_t^{\mathrm{LM}}\bigr],
\end{equation}
with \([\,\cdot\,;\,\cdot\,]\) denoting vector concatenation. Finally, the output hypothesis is

\begin{equation}
  y_t = \mathrm{softmax}\bigl(\mathrm{DNN}_{\theta}(s_t^{\mathrm{DF}})\bigr),
\end{equation}
where \(\mathrm{DNN}_{\theta}\) is a small feed‑forward network whose weights \(\theta\) are also adapted \cite{Gulcehre2015,toshniwal2018comparison}.

To benefit from LLMs in this context, Mittal et al. \cite{mittal2024salsa} introduced a novel method that, to some extent, resembles deep fusion paradigm. This strategy tightly integrates a pre-trained ASR model with an LLM to enhance transcription accuracy, especially for low-resource languages. Speedy ASR-LLM Synchronous Aggregation (SALSA) synchronously couples the decoder layers of the ASR model (i.e., Whisper \cite{radford2023robust}) with a subset of LLM decoding layers (LLaMA2 7B and 13B) through the training of lightweight projection layers across a number of ASR decoding states. To address the challenge of differing tokenization between the ASR and LLM, the method employs cascading tokenization, where the LLM continues generating the transcript until a valid piece of text that is also valid for the ASR tokenizer is produced, ensuring seamless interaction between the models. This text is then tokenized and fed into the ASR decoder for generation of next states.

While deep fusion introduced a more principled approach to integrating a pre‑trained language model into ASR through a gating mechanism, its inherent limitations have resulted in modest performance gains and, consequently, limited methodological extensions in recent years. Deep fusion estimates the gating parameter by only looking at the LM states, and the estimated gating is not fine-grained, both of which limit its performance. Therefore, most subsequent works have shifted toward cold and shallow fusion paradigms, which offer greater flexibility in terms of training and decoding, respectively. Nevertheless, deep fusion retains relevance as a foundational method for evaluating LM integration techniques within the evolving landscape of ASR research, particularly in specialized contexts like low-resource language scenarios \cite{zhang2025end}.

\begin{figure}[h]
    
    \centering
    \includegraphics[width=0.7\textwidth, height=0.28\textwidth]{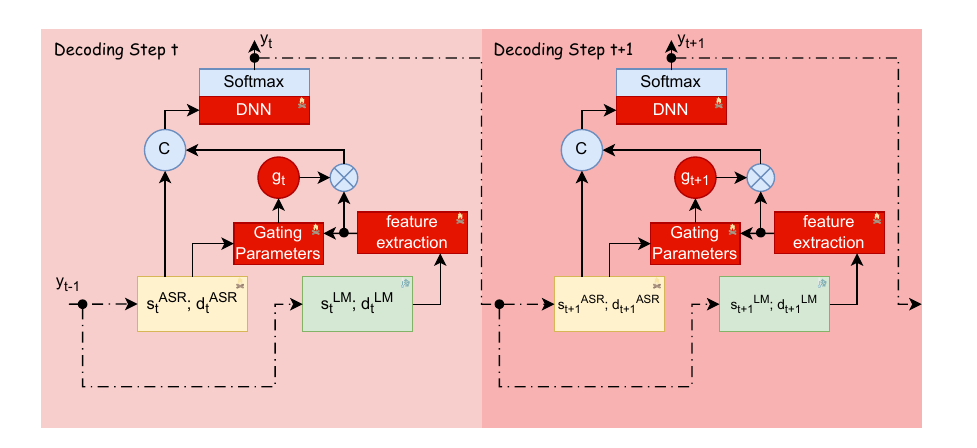}
    \caption{Schematic of Cold Fusion in Two Consecutive Decoding Steps.}
    \label{fig:cold}
\end{figure}

\subsubsection{Cold Fusion}

Based on deep fusion and to mitigate its limitations, Sriram et al.~\cite{sriram2018cold} proposed cold fusion, illustrated in Figure \ref{fig:cold}, that trains an encoder-decoder ASR alongside a fixed pre-trained LM by: projecting the LM's logits through a DNN to generate features \( h_t^{\text{LM}} \) (see Equation \ref{cf-lm1}), computing a fine-grained gate \( g_t \) using both ASR and LM information (see Equation \ref{cf-lm2}), fusing the decoder's hidden state \( s_t \) with the gated LM features \( (g_t \odot h_t^{\text{LM}}) \) into a fused state \( s_t^{CF} \) (see Equation \ref{cf-lm3}), and processing this fused state through another DNN before softmax prediction (see Equation \ref{cf-lm4}). Unlike shallow fusion that blends LM and decoder scores only during the inference and deep fusion that performs partial retraining on the gating mechanism to learn a scalar gating parameter, cold fusion requires full retraining of the ASR decoder (or encoder-decoder). It utilizes both ASR and LM states to predict a fine-grained gating vector to deeply integrate the LM’s linguistic knowledge into the ASR model.

\begin{equation}
    \label{cf-lm1}
  h_t^{\text{LM}} = \mathrm{DNN}_{\theta_1}\bigl(d_t^{\text{LM}}\bigr),
\end{equation}

\begin{equation}
\label{cf-lm2}
  g_t = \sigma\bigl(W_g\bigl[s_t;h_t^{\text{LM}}\bigr] + b_g\bigr)
\end{equation}

\begin{equation}
\label{cf-lm3}
  s_t^{CF} = \bigl[s_t;g_t \odot h_t^{\text{LM}}\bigr]
\end{equation}

\begin{equation}
\label{cf-lm4}
  y_t = \mathrm{softmax}\bigl(\mathrm{DNN}_{\theta_2}(s_t^{CF})\bigr)
\end{equation}

To further improve cold fusion, Cho et al.~\cite{Cho2019} introduce three memory‑control fusion schemes that integrate a frozen RNN-LM into an encoder–decoder ASR system.
The first scheme updates the decoder’s Long Short-Term Memory (LSTM)~\cite{hochreiter1997long} cell state at each step by adding a gated LM output to the original cell memory.
The second scheme combines this cell‑control update with the standard cold fusion method \cite{sriram2018cold} by also gating and concatenating LM information into the hidden state before prediction.
The third scheme extends these dual updates with an affine gating mechanism that fuses LM features into both cell and hidden states prior to the softmax layer, allowing the model to capture more linguistic information when predicting the next hidden and cell states.
Moreover, In \cite{Kim2021}, Kim et al. extend prior cold fusion strategy by enabling an external NLM to be tightly integrated into an RNN-T both during training and inference without adding any extra algorithmic latency. They introduce lightweight gating and fusion modules that project LM logits into a fusion space and merge them with the transducer’s prediction network in a plug-and-play fashion, allowing any pre-trained NLM to be swapped in without re-training the core ASR model.

\subsection{Rescoring}
\label{method-rescoring}
Rescoring is another ASR refinement technique by re-evaluating the initial set of partial or full candidate transcriptions, known as hypotheses, generated by the ASR system in the first-pass decoding. 
During this pass, ASR systems typically produce a lattice \(L\) of possible word sequences or further an n-best list based on acoustics, weak linguistic knowledge, and using a decoding scheme, like in \cite{chow1989n,mohri2002efficient,tran1996word}. 
Rescoring involves re-evaluating these hypotheses using a rich source of linguistics (i.e., LM) to find the most plausible transcription, thereby improving overall recognition accuracy.
The core idea is to combine acoustic model (AM) probabilities \(p_{AM}\) or confidence score in seq2seq ASR models with probabilities of linguistic-rich external LM \(p_{LM}^{External}\) and select the most probable candidate \(y^{*}\). 

% Presenting mathematical formulation
The mathematical foundation of rescoring is rooted in Bayes' theorem, where the goal is to find a sequence of words \( y=(w_0,\dots,w_{T-1})=w_0^{T-1} \) that maximizes the posterior probability \(p(y\mid O)\):
\begin{equation}
\label{eq:rescoring-theorem}
y^* = \argmax_y p(y \mid O) = \argmax_y \frac{p(O \mid y) \cdot p(y)}{p(O)}
\end{equation}
where \(O\) stands for acoustic features. Since \( p(O) \) is constant across hypotheses, Equation \ref{eq:rescoring-theorem} simplifies to:
\begin{equation}
y^* = \argmax_y p(O \mid y) \cdot p(y).
\end{equation}
In practice, to handle numerical stability and incorporate tuning parameters, rescoring often uses a log-linear interpolation:
\begin{equation}
\label{eq:rescoring-main}
S(y) = (1-\lambda)\log p(O \mid y) + \lambda \log p(y) + \mu |y|
\end{equation}
Here, \( p_{AM}=p(O \mid y) \) is the acoustic likelihood (representing how well the audio matches the hypothesis) or confidence score in seq2seq ASR, \( p_{LM}^{External}=P(y) \) is the prior probability of hypothesis $y$ from an external, advanced source of linguistics, \( \lambda \) is a tunable hyper-parameter, \( \mu \) is the word insertion penalty (a scalar that penalizes or encourages longer/shorter hypotheses to control verbosity), and \( |y| \) is length of the hypothesis. Finally, the rescoring process that aims to find the optimal transcript $y^*$ that both matches the acoustic information and is linguistically plausible, is simply formulated as follows:
\begin{equation}
\label{eq:max-s_y}
    y^* = \argmax_yS(y).
\end{equation}

Rescoring studies can generally be categorized into two branches: (i) first-pass rescoring, which improves the initial acoustic likelihood or arc/edge weights within the lattice, and (ii) second-pass rescoring, which re-evaluates the n-best hypotheses after they have been fully generated.

In the first-pass rescoring, an external source of knowledge is applied to the lattice or word graph \cite{ortmanns1997word}, produced by the initial decoding of the ASR system, to adjust arc weights (acoustic scores) and then extract an improved n-best list or final transcription. The intuition is that lattices and word graphs retain rich acoustic alternatives and include many more hypotheses compared to the n-best list. However, these alternatives are less linguistic-aware and assigned scores are consequently less plausible. So adding linguistic information at this stage can promote candidate extraction. Notably, because lattice rescoring must handle a large number of candidate sequences, methods applied at this stage must be fast and sufficiently optimized to be integrated into the real-time applications.

The second-pass approach, which is more studied recently, suggests that including a rich and efficient rescoring model in the first pass is often impractical for real-time scenarios. Plus, seq2seq ASR models do not provide any first-pass, time-aligned lattice to be re-evaluated. Instead, it applies rescoring to the fully generated n-best list from the ASR system. This approach allows for the flexible integration of advanced LMs or even LLMs during rescoring without compromising the efficiency of the initial decoding, particularly when used in the two-pass manner that we will discuss later. Still, second-pass rescoring can be slow and particularly is limited by the weaker scoring method in the first pass \cite{deoras2011fast}. 
Therefore, the first pass affords a broader acoustic-aware search space while constraining computational complexity, whereas second-pass rescoring permits increased computations at the cost of diminished acoustic awareness. So, algorithms in the first pass mostly seek an optimized way to efficiently include linguistics, while in the second-pass the usefulness of acoustic and phonetic information is being evaluated. 
In the remainder of this section, we examine the challenges and describe the innovative methodologies proposed for both rescoring scenarios.

N-grams are the primary LMs to be used for rescoring, that assume the probability of each word \(w_i\) in a sequence \(y=w_0^{T-1}\) depends on mere previous \(n-1\) words. Hence, the probability of each hypothesis is measured in the following manner:

\begin{equation}
\label{eq:ngram-rescoring}
    p(y) = \prod_{w_{i}}p\bigl(w_i \mid w_{i-1}, w_{i-2}, \dots, w_{i-n+1}\bigr).
\end{equation}

% first pass and second pass rescoring using N-GRAMs
In an early study, Tran et al. \cite{tran1996word} proposed first-pass rescoring with n-gram LMs in a two-stage algorithm primarily designed for n-best search from the word graph and highlighted the importance of rescoring. The authors recommend rescoring of the word graph at the phrase level with bigram and trigram LMs before selecting the n-best hypotheses using a tree-organized n-best search. Moreover, they evaluate second-pass rescoring of the extracted n-best list using a four-gram (4-gram) LM and report reductions in WER and improvements in perplexity resulting from the rescoring steps.

% improving lattice rescoring based on N-GRAMs
Murthy et al.~\cite{murthy2024initial} proposed a two-stage method to improve efficiency of first-pass lattice rescoring for low-resource ASR using n-grams by generating an advanced initial lattice. They observe that the weakly trained baseline LM, used for initial decoding of ASR, often omits many out-of-training (OOT) words, which makes straightforward lattice rescoring only marginally effective. To mitigate this, they first minimally augment the baseline LM with unigram counts of OOT words using count-merging interpolation~\cite{pusateri2019connecting} to produce an OOT-inclusive lattice. Second, they rescore this advanced lattice with another augmented n-gram LM. This method substantially enhances performance of lattice rescoring and improves OOT recovery while reducing WER without degrading in-vocabulary recognition performance.

% positive points on DNN-based methods
Methods based on n-grams have limited performance. This limitation arises because, according to Equation~\ref{eq:ngram-rescoring}, the context length they can capture is inherently restricted to \(n-1\) previous words. As a result, they handle pivotal long-range dependencies in language poorly, as discussed in \cite{church2011pendulum}. With the advent of DNNs and NLMs~\cite{bengio2003neural}, especially recurrent \cite{mikolov2010recurrent} and attention-based \cite{vaswani2017attention} architectures, and their superior ability to capture longer dependencies, DNNs were increasingly utilized for rescoring scenarios. Probability of a sequence $y=w_{0}^{T-1}$ with DNNs is measured by:

\begin{equation}
\label{eq:dnn-rescoring}
    p(y) = \prod_{w_i}p(w_i\mid c_i)
\end{equation}
where \(c_i\) denotes the context of token \(w_i\) in \(y=w_0^{T-1}\), which is no longer limited to a few preceding words. For unidirectional architectures, \(c_i\) is the history of preceding tokens \(w_{<i}=w_0^{i-1}\); for bidirectional models (mainly used in second-pass rescoring), \(c_i\) can be the context of the entire sequence (all tokens in the hypothesis). 

% challenges of DNN-based methods
% problem of LATENCY
DNNs also face challenges in performing an efficient rescoring process. 
Firstly, a key and common challenge is that realizing the full capability of these architectures requires deep networks and consequently, a large amount of training data \cite{huang2019empirical}. Conversely, using large DNNs increases the latency of the overall ASR and refinement system. This increase may render such systems impractical in many real-world scenarios. This is particularly true in first-pass rescoring, where the search space of candidate hypotheses is too large and the exact computation of all states using NLMs is not tractable. Moreover, in DNN-based models in order to measure the probability a of each token given its context \(p(w_i\mid c_i)\), a \(softmax\) function is required to be calculated, shown in Equation~\ref{eq:dnn-rescoring-softmax}.

\begin{equation}
\label{eq:dnn-rescoring-softmax}
p(w_i\mid c_i)
= \frac{\exp\big(z_{w_{i}}(c_i)\big)}
{\displaystyle\sum_{v\in\mathcal{V}}\exp\big(z_v(c_i)\big)}
\end{equation}
where \(z_{w_{i}}(c_i)\) and \(\mathcal{V}\) are logit of the DNN corresponding token \(w_i\) within context \(c_i\) and the full vocabulary, respectively.
According to Equation \ref{eq:dnn-rescoring-softmax}, \(softmax\) calculation is exposed to a summation over the \(\mathcal{V}\) that is computationally expensive. Therefore, inherent practical costs of probability estimation also exist.

Several studies have proposed DNN-based rescoring schemes to leverage the strengths of DNNs while mitigating their limitations. Below, we first review first-pass techniques that either rescore the acoustic scores on the lattice or generate it with linguistically informed scores. Next, second-pass methods that re-rank n-best lists or select the top hypothesis directly will be discussed.

% lattice into sub-lattice for lattice rescoring
In an early work on using DNNs, Deoras et al.~\cite{deoras2011fast} proposed limiting the search space on the lattice to use RNNLMs~\cite{mikolov2010recurrent} for first-pass rescoring. They suggested time-based splitting of the lattice \(L\) into \(C\) self-contained ``islands of confusability'' and iterative rescoring of each sublattice \(L_i\) while keeping the surrounding context fixed. Denoting the optimal path in sublattice \(L_i\) by \(\pi_i^*\) and its corresponding word sequence by \(W[\pi_i^*]\), the final decoded hypothesis is achieved by the concatenation of optimal word sequences
\(
y^* = W[\pi_1^*]\cdot\ldots\cdot W[\pi_C^*].
\)
\textit{Entropy-based pruning} is applied to skip sublattices with low uncertainty and further reduce computation. The method computes the posterior entropy of each sublattice \(\mathcal{L}_i\) as:
\begin{equation}
\label{eq:entropy-pruning}
    H(\mathcal{L}_{i})=-\sum_{\pi\in\mathcal{L}_{i}} p(\pi)\log p(\pi).
\end{equation}
If \(H\) is below a tuned threshold, the method skips or aggressively prunes that sublattice (e.g., keeping only the top paths), thereby avoiding expensive RNNLM rescoring on high-confidence (low-entropy) regions. This strategy enables RNNLMs to rescore lattices efficiently. While dramatically shrinking the search space of evaluation, the authors report speeding up and relative WER reduction compared to second-pass n-best rescoring.

%push-forward
In another approach, Auli et al.~\cite{auli2013joint} proposed limiting the number of LSTM~\cite{hochreiter1997long} states retained at each lattice node. They introduced the \textit{push-forward} algorithm, which propagates the RNN states along the lattice arcs by extending each state with the word on the arc and, at every node, retains the top-\(k\) states ranked by cumulative acoustic and RNNLM scores. In the special case where \(k=1\), the original lattice structure is preserved, yielding a particularly efficient rescoring approximation.

Based on the \textit{push-forward} method, Kumar et al. \cite{kumar2017lattice} developed the \textit{arc-beam} algorithm for lattice rescoring, which allows each outgoing arc from each node on the lattice to independently select the best predecessor LSTM state to score its word; unlike the \textit{push-forward} which retains the top-k LSTM states at each node and, when \(k=1\), keeps a single best state per node for rescoring of all outgoing arcs. The authors also employ the sampled softmax technique, which limits the expensive summation over the vocabulary \(\mathcal{V}\) to a uniformly selected subset to reduce the cost of output layer. This method yields relative improvement in WER compared to lattice rescoring based on n-gram LMs.

In another study on using RNN architectures for first-pass lattice rescoring, Liu et al.~\cite{liu2016two}, motivated by the idea of sharing context across similar hypotheses to improve efficiency, developed two context-clustering schemes for RNNLM-based first-pass rescoring: \textit{n-gram-based history clustering} and \textit{history-vector-based clustering}. In both methods, a hash-table-based cache is first implemented that stores RNNLM states associated with a set of distinct contexts \(\tilde{c}_{j-1}=(\tilde{w}_{0},\dots,\tilde{w}_{j-1})\) together with encodings of their context histories \(\Psi(\tilde{c}_{j-1})\). 
Two algorithms are then used to find context matches from the cache during rescoring and to approximate context vectors so that \(\Psi(c_{i-1}) \approx \Psi(\tilde{c}_{j-1})\). In both schemes the key is to find matches that minimize the Kullback–Leibler (KL) divergence between the distributions \(p(\cdot\mid c_{i-1})\) and \(p(\cdot\mid \tilde{c}_{j-1})\), i.e. minimize \(D_{\mathrm{KL}}\big(p(\cdot\mid c_{i-1}) \,\|\, p(\cdot\mid \tilde{c}_{j-1})\big)\).
In the n-gram-based method, the authors find context-history matches based on the similarity of the most recent \(N-1\) words, comparing \(w_{i-N+1}^{\,i-1}=(w_{i-N+1},\dots,w_{i-1})\) with \(\tilde{w}_{j-N+1}^{\,j-1}\). The intuition is that, in RNNLMs, the impact of distant words on the context vector gradually diminishes, so it is reasonable to estimate the context from the most recent \(N-1\) words. 
In the history-vector-based approach, a context match is found by requiring the identity of the most recent preceding word (\(w_{i-1}=\tilde{w}_{j-1}\)) and by measuring the Euclidean distance \(D\) between the corresponding history vectors \(\Psi(c_{i-2})\) and \(\Psi(\tilde{c}_{j-2})\). Thus this strategy effectively considers the full preceding history via the history vectors while giving special weight to the most recent word. 
Because multiple hypotheses can satisfy these criteria, the authors also introduce a lattice-node score-ranking mechanism to reduce sensitivity to traversal order during rescoring. They show that these methods, while more efficient, produce 1-best outputs comparable to second-pass rescoring of large n-best lists.

% GENERATE LATTICE FROM TRACEBACK
Sundermeyer et al.~\cite{sundermeyer2014lattice} introduced two efficient approximations for generating the rescored lattice from traceback~\cite{ney1997progress}: \textit{replacement} and \textit{traceback lattice}. The \textit{replacement} approximation retains the original lattice topology and efficiently assigns LSTM probabilities by following traceback paths and overwriting the LM scores on the corresponding arcs (i.e., tracing best hypotheses back from the final node and updating only the visited arcs while leaving previously updated arcs unchanged). In contrast, the \textit{traceback lattice} approximation explicitly rebuilds the lattice from the traceback tree for LSTM-based rescoring. So, the rescored lattice’s overall best path matches the best path found by lattice decoding. During this process, pruned paths of the lattice are recovered by recombining each pruned traceback path with an appropriate surviving traceback path at the same node.
% GENERATE BETTER SCORED LATTICE 
Xu et al.~\cite{xu2018pruned} also propose generating an advanced lattice by composing linguistic knowledge in the initial decoding. The authors propose constructing a Finite State Transducer (FST) \(B\) such that, when composed with the initial decoder \(A\), it removes the original n-gram (weak LM) cost and inserts the RNNLM cost in a single pass, yielding \(C = A \circ B\), where \(\circ\) denotes the composition operation \cite{mohri2002weighted}. To make RNNLM-based rescoring tractable, rather than performing full composition (which is exponentially growing), the authors introduce a \textit{pruned composition} that treats expansion at the arc level with a priority queue, utilizing a heuristic similar to A* search~\cite{hart1968formal} to prioritize the expansion of ``probably best`` paths and reducing the search space. This pruned composition yields better WER while reducing runtime multiple times speedups versus the \textit{n-gram-based history clustering} approximation rescoring~\cite{liu2016two}. 

% COMPACT RESORING ATTENTION-BASED
So far, several key contributions to first-pass rescoring have been discussed. We now move on to studies that address the second-pass scenario and employ NLMs for this purpose. Huang et al.~\cite{huang2019empirical} address the latency of large DNNs and the high cost of softmax computations by demonstrating that a compact transformer can be made effective for n-best rescoring through a careful combination of design choices. To this end, the authors picked subword tokenization (Byte-Pair Encoding (BPE) \cite{sennrich2015neural}) that interpolates word- and character-level representations to shorten the size of the vocabulary \( |\mathcal{V}| \), and adopted the \textit{adaptive softmax} \cite{joulin2017efficient}, which allocates larger capacity to frequent tokens, both to mitigate the costly \textit{softmax} calculation. For training, they incorporated a high-capacity teacher model, and knowledge distillation (KD) \cite{hinton2015distilling}, to transfer teacher's behavior to a small student model by optimizing:

\begin{equation}
\mathcal{L}_{Total}
=
\mathcal{L}_{\mathrm{KD}}
+
\alpha\,\mathcal{L}_{\mathrm{CE}}.
\end{equation}
where \(\mathcal{L}_{\mathrm{KD}}\) is the KL divergence loss, computed from the small student model predictions and soft labels of the teacher model, and \(\alpha\) is a tunable hyper-parameter to balance CE and KD losses. Empirically, the paper shows that these techniques together yield compact transformer rescoring models that deliver substantial ASR gains. 

Guo et al.~\cite{guo2019spelling} proposed another transformer-based rescoring method for E2E ASR. To expand the search space of candidates for rescoring, this method first adopts a trained encoder-decoder transformer that generates $M$ refined sequences for each hypothesis in the n-best list. Then, it uses another external LM to assign a probability to each of these refined sequences. Finally, it reranks the $M \times N$ hypotheses using a combination of ASR, encoder-decoder refiner, and external LM scores, and selects the optimal sequence as the final transcript.

% use of ACOUSTIC For RESCORING
In another study for second-pass rescoring and with the primary objective of managing low data availability, Gandhe et al. \cite{gandhe2020audio} propose E2E training of a DNN to rescore the n-best hypotheses. This is a simpler task to learn and requires less data compared to full E2E ASR training. The RNNLM in their work directly learns to rescore n-best hypotheses while attending to acoustic features \(p_{LM}^{External}(y \mid O)\). For this, they apply minimum word error (MWE) training (see Section \ref{sec:mwe}) and, to stabilize the training procedure, interpolate the MWE objective $\mathcal{L}_{MWE}$ with the CE loss $\mathcal{L}_{CE}$, weighted by a hyperparameter $\lambda$:

\begin{equation}
\label{eq:rescoring-mwe}
    \mathcal{L}_{Total} = \mathcal{L}_{MWE} + \lambda\,\mathcal{L}_{CE}.
\end{equation}

Since there is no direct alignment between each word $w_i$ and the acoustic features \(O\), the LSTM-based RNNLM is used to learn this dependence and

\begin{equation}
    p\bigl(w_i\mid w_{<i};O\bigr) = \operatorname{softmax}\bigl(\mathrm{LSTM}(w_{<i},O)\bigr).
\end{equation}

Therefore, the rescoring function (Equation \ref{eq:rescoring-main}) becomes \(S(y) = p_{\mathrm{LM}}^{\mathrm{External}}(y \mid O)\) where \(y \in \{y_i\}_{i=0}^{N-1}\). Gandhe et al. report relative improvement in WER compared to E2E training of low-resource ASR systems.

% PAIR-WISE RE-RANKING (DISCRIMINATIVE)
Ogawa et al.~\cite{ogawa2018rescoring} also introduce a second-pass discriminative pairwise framework that directly compares hypotheses to predict which one has a lower WER. It consists of an LSTM-based encoder that converts each hypothesis into a fixed-length representation and a feedforward binary classifier that estimates the probability of one hypothesis being better than another. By repeatedly applying these one-versus-one comparisons, the model reranks the n-best list to minimize expected WER. This pairwise strategy allows the model to learn relative quality distinctions rather than absolute scores, making it effective for discriminative rescoring. Experimental results show that this model outperforms the LSTM baseline that assigns a probability to each hypothesis, achieving about a \(10\)\% relative WER reduction.

Despite providing improvements through advanced LMs and being less computationally limited, second-pass rescoring methods are not applicable to real-time settings. To address this, several studies have proposed two-pass rescoring schemes that enable second-pass rescoring in real-time applications.
In two-pass rescoring, a streaming ASR model (e.g., RNN-T) generates hypotheses, which are then rescored by a non-streaming second-pass model that has access to the full utterance and the partial n-best hypotheses, selecting the best output. Sainath et al.~\cite{sainath2019two} proposed an early approach using a LAS (Listen, Attend and Spell) second-pass module~\cite{chan2015listenattendspell}. Their training procedure is as follows: first, train an RNN-T model; next, take the encoder from that RNN-T, freeze it, and train a LAS decoder on top of the frozen encoder; finally, perform deep fine-tuning by jointly training the shared encoder and both decoders with a combined loss. The resulting system can then be further optimized with MWE training using RNN-T hypotheses. At inference, the RNN-T emits an n-best list which LAS then re-scores. 
In a similar approach, Li et al.~\cite{li2020parallel} adopt a transformer-based architecture for second-pass rescoring of the initial hypotheses produced by a streaming RNN-T model. An additional encoder processes embeddings of the RNN-T encoder, and a transformer decoder is then used to rescore the top hypotheses generated by the RNN-T on the top of the additional encoder. During training, when the RNN-T is trained and frozen, the second-pass model is first trained with CE loss to predict target transcriptions and is then fine-tuned using a MWE objective for discriminative rescoring of the RNN-T hypotheses. 

Hu et al.~\cite{hu2020deliberation} follow the same rescoring pipeline but replace the classic LAS rescoring with a deliberation network~\cite{xia2017deliberation}. In this approach, each n-best hypothesis is encoded bidirectionally, the resulting encodings are concatenated, and these concatenated hypothesis representations, together with the acoustic encoder outputs, are attended to by the LAS decoder to compute log-likelihoods and re-rank the n-best list. This fusion of textual context and acoustic evidence improves disambiguation for proper names, URLs, numerics, and other challenging cases. In a subsequent work, Hu et al.~\cite{hu2021transformer} replace the LSTM stack with transformer decoder layers. The transformer rescorer jointly attends to encoded audio and n-best hypotheses, augments each wordpiece with a learned hypothesis-order embedding, and produces audio and text context vectors. A merger layer then combines these into a single context vector matching the self-attention dimensionality. This design enables parallel token scoring, stronger multi-source fusion, and avoids a costly second beam search.
In a subsequent multilingual extension, Hu et al.~\cite{Hu2022Scaling} again adopt the same deliberation-style rescoring framework but replace the monolingual hypothesis encoder with a multilingual text encoder. By encoding n-best hypotheses in a shared multilingual representation space, the model enabled unified rescoring across multiple languages.

With advances in general-purpose bidirectional LMs (e.g., BERT) trained on large text corpora, many studies have incorporated these models into both discriminative and non-discriminative rescoring for ASR systems (see \cite{shivakumar2023discriminative} for a detailed discussion), leveraging their pre-trained knowledge to perform rescoring without extensive retraining. 
An important branch of studies involve employing masked language models (MLMs) that leverage bidirectional contextual information to improve transcription understanding.
Specifically, Shin et al.~\cite{shin2019effective} introduce a bidirectional self-attention network language model (biSANLM), a modified BERT that omits masking during inference and is trained solely on masked word prediction with adjustments such as single-sentence inputs, consistent \texttt{[MASK]} replacements, and a limit of up to four masked tokens per instance. It scores sentences by sequentially masking each word in the hypothesis and computing the log-likelihood of the original word using both left and right contexts. Summing these values yields a pseudo-probability for rescoring, allowing interactions across the entire sentence via the \texttt{[MASK]} token without trivializing the task.

Building upon the previous idea, Salazar et al.~\cite{salazar2020masked} formalize pseudo-log-likelihood (PLL) scoring with an MLM like BERT to assess linguistic fluency in a bidirectional manner, defining PLL for a sentence $y = w_{0}^{T-1}$ as:

\begin{equation}
\mathrm{PLL}(y) = \sum_{i=0}^{T-1} \log p(w_i \mid y_{\setminus {w_i}}),
\end{equation}

where $y_{ \setminus {w_i}}$ denotes the sentence while $w_i$ is being masked. The rescoring interpolates the ASR score $\log p(y\mid O)$ (from acoustic input $O$) with the PLL-based LM score via Equation \ref{eq:rescoring-main} where $p(y) \approx \mathrm{PLL}(y) / |y|$ or similar normalization. The authors also introduce a maskless student model, fine-tuned on the \texttt{[CLS]} token representation, to approximate PLL in a single efficient pass, decoupling ASR and LM for flexibility and reducing computational cost from $\mathcal{O}(|y|)$ inferences to $\mathcal{O}(1)$.
Extending this line of research, Udagawa et al.~\cite{udagawa2022effect} conducted rescoring experiments using bidirectional language models such as BERT and RoBERTa~\cite{liu2019roberta} on a more competitive ASR system, the Conformer-Transducer, compared to the models used in~\cite{shin2019effective} and~\cite{salazar2020masked}. They showed that bidirectional LMs improved ASR performance, whereas a unidirectional LM failed to do so.
Building on the approach of Salazar et al.~\cite{salazar2020masked}, RescoreBERT~\cite{xu2022rescorebert} extends PLL-based rescoring by incorporating discriminative training of ASR hypotheses using scores derived from the \texttt{[CLS]} token representation. To achieve this, they applied the MWE loss that minimizes the expected WER across n-best hypotheses. In addition, they proposed a novel Matching Word Error Distribution (MWED) loss, which aligns the model’s predicted distribution over hypotheses with the true word-error distribution, effectively optimizing the Kullback–Leibler divergence between the two.

Chiu and Chen~\cite{chiu2021innovative} proposed PBERT, which employs discriminative training with BERT for hypothesis rescoring but without relying on pseudo-log-likelihood (PLL) computations. They use the \texttt{[CLS]} embeddings of hypotheses followed by a fully connected network and a softmax layer to select the oracle hypothesis with the minimum WER. Furthermore, they incorporate Probabilistic Latent Semantic Analysis (PLSA)~\cite{hofmann2001unsupervised} to capture task-specific global topic information in an unsupervised manner, enhancing PBERT for n-best hypothesis reranking and referring to this extended model as TPBERT.

In recent years, the emergence of LLMs has drawn attention in ASR research. This is primarily due to their exceptional ability to generalize across tasks and domains through in-context or few-shot learning. Additionally, LLMs exhibit far stronger linguistic understanding compared to ASR systems trained with limited linguistic knowledge. As a result, researchers are increasingly using them for ASR rescoring. In a pioneering study on using LLMs for rescoring, Ma et al. \cite{ma2023can} propose using GPT \cite{achiam2023gpt}, a generative LLM, for ASR refinement in two modes: unconstrained and constrained. In the constrained mode, which resemble the second-pass rescoring, GPT is simply instructed to select the most probable hypothesis from the n-best list as the final transcript. In unconstrained mode (which is placed in Section \ref{sec:llm-based}) the LLM is instructed to rewrite a correct transcript based on the initial hypotheses. The authors report that the proposed system can improve ASR accuracy in both modes. 
Tur et al. \cite{tur2024progres} extended the approach of \cite{ma2023can} and introduced PROmpted Generative REScoring (PROGRES). In this method, an LLM is first instructed to produce an additional transcript conditioned on the n-best list of pre-trained ASR system. The LLM may either select an existing hypothesis or generate a new one from scratch; this yields an extended candidate list of \(n+1\) hypotheses. The intuition is that hypotheses in the n-best list are largely driven by acoustic evidence and thus contain limited linguistic information, which LLMs, as rich sources of language knowledge, can provide. Next, the extended list is rescored using another LLM (i.e. Llama-3), and the hypothesis with the highest score is selected as the final transcription (ASR score for the generated hypothesis is set equal to the highest probability found in the initial n-best list). The authors report that extending the n-best list improves the rescoring performance, and that interpolating ASR and LLM scores outperforms either LLM-only or ASR-only scoring.

More recently, Shivakumar et al.~\cite{shivakumar2025speech} investigated speech-text foundation models, LLMs trained on both modalities, for second-pass ASR rescoring. The authors evaluate two model sizes: a 330M-parameter OPT-like architecture~\cite{zhang2022opt} and a 7B-parameter LLaMA-based model~\cite{touvron2023llama}. Both models were pre-trained on large corpora containing unlabeled speech, text, and parallel speech–text data. Speech representations were extracted using HuBERT~\cite{hsu2021hubert} and clustered with k-means, while text was tokenized using SentencePiece tokenizer~\cite{kudo2018sentencepiece}. As a multimodal rescoring technique, the approach scores ASR n-best hypotheses by conditioning on concatenated speech and text token sequences as:

\begin{equation}
\label{eq:foundation-model-rescoring}
p_{FM}(Z) \;=\; \prod_{i=0}^{T-1} p\big(z_i \mid z_{<i}\big)
\end{equation}
where \(p_{FM}\) is the probability of the foundation model assigned to \(Z=(z_0,\ldots,z_{T-1})\) that is the multimodal sequence, \(z_i \in \mathcal{V}_{\text{text}} \cup \mathcal{V}_{\text{speech}}\), and \(\mathcal{V}_{\text{text}}\) and \(\mathcal{V}_{\text{speech}}\) denote the text and speech vocabularies, respectively. The authors investigate the effect of modality order and find that speech-first configurations (audio tokens preceding text tokens) yield superior performance compared with text-first and text-only baselines. Additionally, the study evaluates discriminative rescoring via MWE fine-tuning and reports further WER reductions, consistent with prior findings by \cite{shivakumar2023discriminative}.

\subsection{Correction}
\label{method-correction}

Unlike other ASR refinement strategies, such as rescoring methods that merely select from existing hypotheses or fusion techniques that adjust the first-pass output, the correction category can generate entirely new transcriptions and often expands the vocabulary. These methods typically rely on one of the following components:

\begin{itemize}[noitemsep]
    \item Rule-based and n-gram.
    \item NLMs.
    \item LLMs.
    \item Retrieval mechanisms that leverage external knowledge bases.
\end{itemize}

In this section, we evaluate the correction methods introduced above and distill the key insights within each category. For clarity, the methods are grouped according to their primary innovations, which generally align with the categories previously outlined. Nevertheless, many approaches integrate multiple components (e.g., error  detection, confidence estimation), which may give rise to diverse processing pipelines or suggest additional categories. However, for consistency, we classify methods based on their core component, as it provides the most coherent basis for comparison. 

We first discuss rule-based and n-gram techniques, which are largely conventional. Then, we examine NLM-based methods, which adopt encoder-only or decoder-inclusive architectures, followed by LLM-based approaches that provide greater modeling capacity. Finally, we review retrieval-based algorithms. This order reflects both the historical progression of research in the field and the increasing sophistication of the underlying methods.

\subsubsection{Rule-based and N-gram}
Early ASR correction efforts relied on rule‐based algorithms and n-gram methods to refine transcripts. However, both approaches exhibit significant limitations. Rule‐based systems struggle with unseen error patterns, since ASR mistakes vary widely across acoustic domains. N-gram models are constrained by their fixed context window. No finite \textit{n}-gram can cover all valid word or character combinations. Choosing a large \textit{n} leads to data sparsity and can even introduce over-corrections. In contrast, a small \textit{n} severely restricts contextual awareness~\cite{church2011pendulum}. Blow, we summarize key contributions in this category.

In an early effort for ASR correction, \cite{Fiscus1997} proposed the Recognizer Output Voting Error Reduction (ROVER) method, which first aligns 1-best hypotheses from multiple ASR systems into the proposed Word Transition Network (WTN), and then adopts a voting scheme to select the best token for each position in the transcript. Recently, \cite{Imai2025Saki} extended ROVER by using n-best hypotheses from a single ASR system instead of 1-best hypotheses from \(N\) different systems, broadening ROVER’s applicability when multiple ASR models are unavailable. 

Based on n-gram LMs, Bassil et al. in \cite{bassil2012asr} developed a three-stage pipeline. It first flags errors in any word absent from a unigram vocabulary, then generates correction candidates using character-level bigrams for detected ASR errors, and finally selects replacements based on word-level 5-gram frequency statistics over the four preceding words in the initial transcript plus the generated candidate.
In another research that to some extent resembles n-gram based techniques, Bassil et al.~\cite{bassil2012post} introduce an innovative post-editing method for correcting ASR errors using Bing's web search engine and its spelling suggestion technology, which leverages the vast indexed database of web content to probabilistically identify and rectify misrecognized words. In this approach, the initial ASR transcript is tokenized into sequences of 6-word blocks \(T=(t_0,\ldots,t_{L-1})\) where \(t_i=(w_{i,0},\ldots,w_{i,5})\), providing sufficient context to disambiguate errors that might not be detectable in isolated words. Each \(t_i\) is submitted as a query to Bing; if the engine detects potential misspellings and returns a suggestion in the form of ``Including results for \(t'_i\)``, where \(t'_i\) is the proposed correction, it replaces \(t_i\) in the transcript; otherwise, \(t_i\) remains unchanged. This sequential process is performed once across all tokens, yielding a corrected transcription by concatenating the validated and modified segments.

Zhou et al.~\cite{zhou2006multi} proposed a multi-stage framework for error detection and correction in Mandarin Large Vocabulary Continuous Speech Recognition (LVCSR) comprising a detection module followed by a six-step correction procedure. The detection module is itself performed in three steps: an utterance classifier that flags potentially erroneous utterances using features based on Generalized Word Posterior Probability (GWPP)~\cite{lo2005generalized} and n-best hypotheses; a word classifier that uses GWPP to locate erroneous words inside those utterances; and an assumption that all characters within words labeled erroneous are candidates for correction. The correction stage is a six-step procedure: it extracts candidate characters from the recognition lattice; forms a search network of utterance hypotheses; segments hypotheses into words; scores each hypothesis with a linearly combined mutual-information and trigram model; ranks the hypotheses and selects top candidates; and finally applies a confidence-dependent threshold to accept corrections to reduce false positives caused by imperfect detection.

D'Haro et al. \cite{DHaro2016Automatic} trained a phrase-based statistical translation model, which constructs a lookup table mapping ASR output phrases to reference phrases, along with the probabilities of each mapping. In addition, an n-gram language model is used to ensure fluency and further correct the output. During inference, the ASR n-best hypotheses are passed through this phrase-based translation model, producing n corrected transcripts. These candidates are then re-ranked using Minimum Bayes Risk~\cite{kumar2004minimum} to select the final corrected transcription.

Zietkiewicz et al. \cite{Zietkiewicz2022} propose a primarily rule-based method for post-editing tokens: they first align ASR hypotheses and reference texts using the Ratcliff–Obershelp similarity algorithm \cite{ratcliff1988pattern} to derive token-level edit operations such as append, delete, and join. These operations are then encoded as rule-based tags applied to tokens in the noisy transcript, forming the core innovation of the approach. A neural model is used only to predict which of these rule-derived tags should be activated, and applying the selected tags directly yields the corrected output.

\subsubsection{NLM-Based} 
\label{sub:nlm-correction}
With the advent of neural networks, particularly following the introduction of transformers \cite{vaswani2017attention}, and advances in computational hardware, the use of NLMs has grown substantially. This development gave rise to NLM-based methods aimed at improving ASR post-editing, complementing the progress achieved in ASR modeling itself. These approaches benefit from the ability to capture much longer dependencies than n-gram LMs, which is pivotal in many NLU tasks, as well as from their enhanced generalization capability.

We categorize NLM-based correction methods into two classes: (1) encoder-based and (2) decoder-inclusive approaches. 
Encoder-based methods focus on context-aware correction of the ASR output, producing candidate corrections for each token position in the initial transcript. Decoder-inclusive methods step further: they adopt a decoder to extend and correct the original ASR hypotheses.
In the following, we summarize key contributions in each category to clarify their distinct scopes.

\begin{enumerate}[leftmargin=1.5em, labelsep=0.5em, itemsep=0.4em]
    \item \textbf{Encoder-Based:} 
\label{par:corr-encoder}
Shifting to NLMs, a distinct line of work leverages the strong contextual modeling capabilities of encoder architectures for ASR error correction. Although encoder models also play vital roles in other classes of approaches, here we focus exclusively on those that exploit only the encoding architecture, offering a lightweight yet effective solution for ASR post-editing.

A common baseline in this paradigm is a pretrained (or further fine-tuned) encoder-based transformer model, most notably BERT \cite{devlin2019bert}, combined with a \(softmax\) prediction head to generate correction candidates at each token position based on contextual information \cite{zhang2020spelling,zhang2021correcting}. Since BERT is trained with a masked language modeling objective analogous to the task of predicting missing or erroneous tokens, it is naturally suited for error-correction applications, as demonstrated in \cite{nanayakkara2022clinical}. The vanilla BERT models, evaluated in \cite{zhang2020spelling,zhang2021correcting}, typically fail to produce large gains in transcription accuracy, mainly due to a mismatch between clean pretraining corpora and noisy domains of ASR transcription. Additionally, because methods based on BERT predict tokens solely from textual context (not from acoustics), they may substitute semantically plausible but phonetically incorrect words, reducing correction quality. We review several methods that guide and improve the encoder-based correction process.

As noted earlier, it is important to consider not only contextual fit but also symbolic and phonological similarity.
To this end, Hong et al. \cite{hong2019faspell} introduced FASPell, which first generates correction candidates using contextual representations of BERT, then uses a decoding algorithm to score each candidate by phonological similarity, visual similarity, and semantic coherence within the sentence.
Cheng et al. \cite{cheng2020spellgcn} proposed SpellGCN, which employs a graph neural network to encode relationships among words based on pronunciation and visual likeness, and then merges those graph‐based features with semantic embeddings of BERT to improve token‐level corrections.
Zhang et al. \cite{zhang2021correcting} presented MLM-Phonetics, which softly masks tokens based on their phonetic embeddings, biasing BERT toward replacements that preserve pronunciation. 

Fan et al. \cite{fan2023boosting} built another BERT-based framework and proposes the dynamic error-scaling mechanism that that enriches each character in the transcript with word-level knowledge derived from a dictionary and pinyin-based matching (Pinyin is the standard and most common system of romanized spelling for transliterating Chinese). For every character, the model retrieves plausible 2-character word candidates and uses a learned char–word attention mechanism to judge how well each candidate fits the sentence context. If a character’s candidates align semantically and phonetically with the surrounding text, the model naturally stabilizes its representation, making it more likely to keep the character (error reduction). If the candidates are noisy, semantically mismatched, or only loosely connected via pinyin, the representation becomes unstable, signaling that the character is likely incorrect (error amplification). These enriched representations feed into a correction decoder that decides whether to copy or modify each character. Overall, the method improves both error detection and correction by allowing word-level and phonetic cues to dynamically guide the model.

To prevent overcorrection and perform targeted corrections, several pipelines introduce an explicit error‐detection module. Zhang et al. \cite{zhang2020spelling} proposed Soft-Masked BERT, coupling a bi-GRU error detector with a soft-masking mechanism that interpolates between the transcript and \texttt{[MASK]} token embeddings, according to detection confidence before feeding them into the correction module. Chen et al. \cite{chen2021integrated} also integrate an error‐detection classifier and strictly masks the erroneous tokens. They then feed these partially masked transcripts into a BERT model to identify correction candidates for the error positions. Finally, the correction is selected by combining phonetic similarity between the erroneous token and the candidate corrections with a trade-off between semantic and phonetic information.

Using speech information can be useful to complement the contextual knowledge of encoder-based models for correction. To this end, Peyghan et al. \cite{peyghan2025cmc} proposed a Cross-Modal Contextualized (CMC) spelling correction method that leverages speech embeddings of WavLM \cite{Chen2021OctWavLM} into a cross-modal correction module to produce speech-aware edits and incorporates a BERT-based detection module to prevent over-corrections. The authors further propose a framework that interpolates embeddings of correction candidates with the original transcript tokens, weighting the interpolation by the detection module's estimated error probability. This approach preserves information from the initial transcript while applying corrections when the detector indicates a high likelihood of error.

\item \textbf{Decoder-Inclusive:} 
\label{method-correction-decoder-inclusive}
Decoder-inclusive methods represent another paradigm within NLM-based error correction, employing either encoder–decoder or decoder-only architectures. By incorporating an explicit decoding module, these approaches extend beyond encoder-based models, enabling correction of a broader range of ASR output errors. Nevertheless, in addition to the challenges discussed in \nameref{par:corr-encoder}, this paradigm introduces specific trade-offs: These methods are more susceptible to over-correction and incur higher computational cost and inference latency due to their generative nature and decoding mechanisms, respectively. To address the susceptibility to over-correction, autoregressive (AR) decoders have traditionally been adopted due to their strong conditional modeling capabilities. Conversely, to alleviate the computational and latency overhead introduced by the decoding process, non-autoregressive (NAR) decoders have been proposed as lighter and faster alternatives. In this work, we introduce decoder-inclusive correction methods according to this taxonomy of decoder architectures, enabling readers to choose the design that best aligns with their practical constraints and performance needs.

\paragraph{Autoregressive (AR) Decoders:} 
% AR decoding
Autoregressive decoders generate each token conditionally based on the previously generated tokens, which allows them to model dependencies effectively and produce high-quality corrections. This conditional generation makes them particularly suitable for accurate ASR error correction. However, this sequential nature also introduces significant computational overhead and latency, making AR decoders less suitable for applications where fast transcription or real-time processing is critical.

In a pioneering work, Tanaka et al. \cite{Tanaka2018NeuralErrorCorrective} introduced Neural Error Correction Language Models (NECLMs) with an encoder-decoder RNN architecture that explicitly learns to correct ASR errors. The model is trained on pairs of erroneous and corrected transcripts, learning to predict the corrected sequence given the noisy ASR output. 
Following this line of research, Zhang et al. \cite{zhang2019investigation} employed a Transformer-based encoder–decoder architecture to correct the outputs of a CTC-based Mandarin ASR system.
In a similar study, Hrinchuk et al. \cite{hrinchuk2020correction} proposed a Transformer-based ASR correction model in which both the encoder and decoder are initialized using BERT parameters. By transferring linguistic knowledge from large-scale pretraining directly into the correction model, they showed that initialization from BERT substantially improves correction accuracy compared to training Transformers from scratch.
Li et al. \cite{li2021boost} enhance transformer-based decoder-inclusive correction by fusing BERT representations into the encoder and adding a copying mechanism \cite{gu2016incorporating} in the decoder. The BERT-fused encoder enriches input features with pretrained linguistic knowledge. In the decoder, a gate mixes two distributions: a normal vocabulary softmax and a copy distribution computed from the cross-attention on the last-layer of the decoder, where the attention weights indicate the importance of each token in the ASR output and thus allow the model to directly reuse tokens that should not be changed. The gate is learned from the decoder’s hidden state, deciding whether to copy or generate, making this mixture a principled and interpretable copy probability.

Although initializing decoder-inclusive architectures using MLM models like BERT provides rich contextual representations, their masked-language-modeling objective is not fully aligned with the task of correcting noisy ASR transcripts. Building on this observation, Zhao et al. \cite{zhao2021bart} showed that initializing an encoder–decoder correction model with BART \cite{lewis2020bart}, whose pretraining objective is to reconstruct clean text from noisy input, outperforms models initialized with RoBERTa \cite{liu2019roberta}.
Yeen et al. \cite{yeen2023learned} introduce a detector–corrector pipeline for ASR error correction using T5 \cite{raffel2020exploring}, which is an encoder–decoder model pretrained with span-corruption denoising that aligns well with the goal of generating clean text from corrupted input. In their framework, token-level error labels are obtained via Levenshtein alignment; incorrect tokens are masked; a generator (MLM) predicts candidate replacements; and a discriminator learns to detect erroneous tokens. At inference time, the generator is removed: the discriminator identifies and masks error tokens in the ASR hypothesis, and the masked hypothesis is then fed to the T5-based model to produce the corrected transcript.

Ma et al. \cite{ma2023nbest} fine-tuned a pretrained T5 model using ASR n-best hypothesis lists as input rather than relying solely on the 1-best output, providing richer contextual clues for correcting errors. The n-best hypotheses are concatenated with special separator tokens and preceded by the prefix “text correction:” to guide the model, leveraging both T5’s multi-task pretraining and the additional context from multiple hypotheses. Their method also incorporates n-best-constrained decoding, which restricts the correction search space to the hypotheses in the n-best list, as well as lattice-constrained decoding, which expands the candidate space by considering paths through the ASR lattice.

Despite the advances achieved by the aforementioned methods, their exclusive reliance on textual information constrains their effectiveness. To address this limitation, Dutta et al.~\cite{dutta2022error} incorporate phoneme sequences generated by a grapheme-to-phoneme (G2P) model~\cite{g2pE2019} alongside the text, and fine-tune a pretrained BART model for ASR error correction.
Building on the idea of leveraging additional speech-related information, Hu et al.~\cite{hu2020deliberation} utilize speech embeddings extracted from the shared encoder of a two-pass ASR system to refine the first-pass transcript. These speech embeddings are further enriched by an additional encoding module, and the first-pass transcript is then fed into a deliberation decoder~\cite{xia2017deliberation}.
Tanaka et al.~\cite{tanaka2021cross} extend this line of work by introducing a transformer-based architecture with cross-modal encoding, which jointly processes textual and acoustic embeddings before decoding. This differs from earlier approaches that encoded each modality separately~\cite{hu2020deliberation}.

More recently, Li et al.~\cite{li2024crossmodal} proposed to use discrete speech units derived from HuBERT~\cite{hsu2021hubert}, and combine them with the 1-best ASR output encoded by a RoBERTa encoder. They apply cross-modal attention between the speech-unit embeddings and the RoBERTa word embeddings to produce acoustically enhanced word embeddings, which are then used via cross-attention in the decoder to generate the corrected transcript.
%More recently, Li et al. \cite{li2024crossmodal} proposed an encoder–decoder correction method that directly extracts speech embeddings using HuBERT \cite{hsu2021hubert} and applies cross-modal attention between these embeddings and text representations to decode the corrected transcript.
Jiang et al.~\cite{jiang2024cross} further explored multimodal correction by incorporating both audio and phoneme modalities alongside text within a contrastive learning framework, encouraging text representations to align with speech and phonetic features. During training, each modality is passed through its own encoder to match the shared encoder–decoder input space, and the model learns through three tasks: text correction, pinyin-to-text conversion, and audio-to-text conversion. Contrastive loss then enforces high cosine similarity for representations of the same utterance across modalities and low similarity for different samples. At inference time, only the ASR transcript is used. However, because the text encoder was aligned with audio and phoneme representations during training, it implicitly benefits from speech and pronunciation information. Kumar et al.~\cite{kumar2023visual}, in a completely different line of work, sought to leverage visual information from videos to guide ASR error correction. They explored two schemes: a modality fusion approach in which text and image modalities are encoded separately and then fused for decoding, and a prompt-based approach that first generates an image caption and then concatenates this caption with the ASR transcript to serve as input to an encoder–decoder correction model.

\paragraph{Non-Autoregressive (NAR) Decoders:} 
% NAR
To address the computational cost and latency associated with AR architectures, non-autoregressive (NAR) decoders have gained significant interest due to their ability to generate tokens in parallel, enabling substantially faster inference. This makes NAR decoders particularly suitable for scenarios where computational overhead is a critical constraint or where real-time correction is required. However, enabling parallel generation requires additional mechanisms to handle variable-length outputs, such as predicting the target length, introducing blank tokens into the vocabulary, or adopting other structural assumptions to approximate the decoding process. While these mechanisms make NAR decoding feasible, inaccuracies in length prediction or limitations in these assumptions can lead to suboptimal corrections, often falling short of the accuracy achieved by AR decoders, despite offering markedly higher decoding speed.

The first NAR error-correction model was introduced by Leng et al.~\cite{leng2021fastcorrect} as \textit{FastCorrect}. It consists of a Transformer encoder, a length predictor that leverages the encoder representations to determine token edits through repetition or deletion, and a NAR decoder that operates on the edited token sequence while attending to the source encodings. During training, they first compute edit-distance-based alignments between source and target sequences. Among candidate alignments, they select those that minimize edits to the source; when multiple alignments satisfy this constraint equally, they prefer those containing higher-frequency target n-grams. These refined alignments provide supervision for training the length predictor.
% NAR Fast-Correct

Building on the previous line of research, Guo et al.~\cite{guo2021global} propose BERT with a Confusion-set guided Replacement Strategy (BERT\_CRS) for encoding Chinese ASR outputs, together with a Global Attention Decoder (GAD) that operates in a NAR manner and assumes a strict one-to-one alignment between input and output characters, allowing only substitution-based corrections and enforcing equal input–output length. The BERT\_CRS encoder replaces characters not only with masked tokens but also with visually and phonologically similar confusion-set characters during training, enabling the model to better learn ASR-like error patterns. The model incorporates two heads for detecting unchanged tokens and one head dedicated to correction, with the combined loss encouraging more accurate error detection and correction. During inference, high-confidence predictions are accepted directly, while low-confidence positions are assigned 
\(k\) candidate characters from the correction head. These candidates, along with the BERT\_CRS encodings, are then fed into the GAD, which models global interactions among candidate tokens across positions to select the most plausible corrections for the uncertain regions.

Shen et al.~\cite{shen2022mask} observed that ASR correction models typically copy around 90\% of input tokens and modify fewer than 10\% of erroneous tokens, and that simply blocking gradients for correct tokens can hurt performance. They proposed randomly masking correct input tokens during training to encourage the model to rely more on contextual cues. They applied this masking strategy and observed improvements in both AR decoding and NAR decoding, in which one-to-one alignment between input and output tokens was assumed to make NAR decoding plausible.

Yang et al.~\cite{yang2022asr} propose a constrained decoding approach for ASR error correction by first identifying only the tokens that require modification. They employ a BERT encoder together with an operation predictor that labels each input token as KEEP, DELETE, or CHANGE. Only tokens labeled CHANGE are passed to a non-autoregressive decoder, which generates their corrected forms in parallel while attending to the encoder representations. The corrected tokens are then inserted back into their original positions, while tokens labeled KEEP remain unchanged and DELETE tokens are removed.

Leveraging n-best ASR hypotheses can provide substantial benefits for error correction, as different hypotheses often contain complementary or partially correct segments that reinforce one another through cross-hypothesis voting. Building on this idea, Leng et al.~\cite{leng2021fastcorrect2} propose \textit{FastCorrect 2}, which extends NAR correction from the top-1 hypothesis to the entire n-best list. The model first aligns all hypotheses using edit distance and phonetic similarity so they can be jointly encoded, enabling the decoder to exploit cross-hypothesis signals. A candidate-selection module predicts the expected correction loss for each hypothesis and selects the one with the lowest predicted loss. The chosen hypothesis is then processed by the length predictor, which determines token repetitions or deletions, and the resulting adjusted sequence is refined by a non-autoregressive decoder. By incorporating multi-hypothesis information, \textit{FastCorrect 2} achieves improved correction accuracy over the original \textit{FastCorrect}~\cite{leng2021fastcorrect}.

\textit{SoftCorrect}~\cite{leng2023softcorrect} also leverages n-best ASR hypotheses but differs from \textit{FastCorrect 2} in two key aspects. Rather than selecting a single hypothesis, it performs token-level selection by choosing a token at each aligned position across the n-best list. As in \textit{FastCorrect 2}, the hypotheses are first aligned; however, the selection is made position-wise instead of hypothesis-wise. Moreover, rather than relying on an explicit length predictor for NAR decoding, \textit{SoftCorrect} adopts CTC-based decoding, enabling natural handling of variable-length outputs. To distinguish reliable from potentially erroneous tokens, the model trains its encoder with an anti-copy regularization loss that introduces a special error token and encourages the encoder to output this token for uncertain or incorrect positions, making the correct–incorrect boundary more explicit. For each aligned position, the encoder’s token probabilities are combined with the original ASR probabilities to compute a confidence score. Tokens with low confidence are marked as likely errors, expanded by repeating them three times, and then passed, together with high-confidence tokens, into a non-autoregressive decoder. Because the decoder is trained with a CTC loss, blank symbols are naturally removed during decoding, producing the final corrected output.

Adding phonetic information can further enhance ASR error correction, because many recognition errors stem from phonetic rather than semantic confusion. Building on this idea, Phonology-based Variable-length Error Correction (PhVEC)~\cite{fang2022non} introduces a NAR detector–corrector framework for Chinese ASR correction that incorporates pinyin tokens to enable variable-length rewriting and produce phonetically similar corrections. The model begins with a detector that performs binary classification to identify incorrect tokens. For each token predicted as erroneous, its corresponding pinyin sequence is appended to that token, and the resulting expanded sequence is passed to a NAR decoder. At the end of decoding, blank tokens are removed and repeated characters are collapsed, since different pinyin sequences may correspond to the same Chinese character.

Futami et al. \cite{futami2022non} propose the Phone-Conditioned Masked Language Model (PC-MLM), a non-autoregressive encoder–decoder model for correcting CTC-based and other greedy-decoding ASR outputs. The encoder receives phonetic sequences, while the decoder takes the ASR output with low-confidence tokens masked and predicts corrected words in parallel. To handle insertion errors, additional masked tokens are inserted during training and the model learns to output a null token for those positions. Phoneme Augmented Transformer for ASR error Correction (PATCorrect) \cite{zhang2023patcorrect} uses two separate encoders to process word-level text and phonetic information, then fuses these representations in multiple ways to generate tag predictions indicating substitution, deletion, or insertion operations for each source word. The predicted tags determine the corrected sequence length, which guides a NAR correction model that integrates text, phoneme, and ASR output encodings. Wang et al.~\cite{wang2022effective} propose a three-stage NAR ASR error-correction system that first reranks n-best hypotheses using BERT, then detects character-level errors and predicts replacement lengths for each erroneous position, and finally masks those positions and applies a self-attention-based correction module that cross-attends to the BERT encodings of the adjusted sequence as well as the textual and phonetic encodings of the selected hypothesis.

Incorporating acoustic information can significantly improve NAR error correction, as it provides complementary signal-level cues that are often missing even in textual or phonetic features. Wang et al. \cite{wang2022effective} use the frame-level alignment output of a first-pass RNN-T model, including blanks, together with acoustic features to enable NAR decoding through a Transformer-based deliberation decoder. The acoustic features produced by the RNN-T encoder are further processed by an additional encoder and then combined with the alignment sequence for Align-Refine \cite{chi2021align} decoding. In this stage, the model applies self-attention over the alignments and cross-attention over the audio features to iteratively refine the alignment for R steps. 

Du et al. \cite{du2022cross} proposed a cross-modal ASR post-processing system that fuses acoustic features from the acoustic model with textual features derived from the 1-best hypothesis into a length predictor, a confidence estimator, and a corrector. First, source tokens may be deleted or repeated based on the predictions of the length predictor. The confidence estimator and corrector then operate on these adjusted tokens in a NAR manner. After this step, the confidence estimator assigns a score to each token, helping to mitigate the widespread problem of over-confident ASR predictions~\cite{kuhn2025evaluating}. Tokens with scores above a fixed threshold are retained, whereas those with lower scores are replaced by the corresponding outputs of the corrector. In the final stage, a voice activity detection module identifies the overlap probability of speakers at the frame level, which is converted to token-level probabilities. These probabilities, along with the probability of the corrected token, allow the system to remove unreliable tokens or entire regions.

Shu et al. \cite{shu2024error} proposed a non-autoregressive error correction system that incorporates a confidence estimator and acoustic information, operating on n-best hypotheses rather than the 1-best. The authors aligned the n-best hypotheses using the edit-distance-based method from FastCorrect2 \cite{leng2021fastcorrect2} to achieve same lengths, incorporating blank tokens to enable variable-length NAR decoding. These aligned candidates are then passed into the confidence estimator module to obtain confidence embeddings. Next, a linear layer learns the weight for each candidate based on its word and confidence embeddings, producing fused word and confidence embeddings accordingly. Finally, an NAR decoder applies self-attention over the fused word embeddings, followed by cross-attention with the acoustic features and then with the fused confidence embeddings; the resulting embeddings are fed into a linear layer to generate the corrected transcription.

\end{enumerate}

\subsubsection{LLM-Based}
\label{sec:llm-based}
The advancement of LLMs and their strong generalization capabilities have rendered them valuable across numerous applications, including ASR post-refinement. In ASR error correction, LLMs have been employed in three primary scenarios: text-only prompting schemes, incorporation of auxiliary information within prompting (e.g., phonetics or speech features), and LLM fine-tuning. In this part, we highlight innovations that adopt LLMs as the core component in ASR correction. Although these methods share similarities with the \nameref{method-correction-decoder-inclusive} approaches, we dedicate a separate part to LLM-based generative correction to underscore its prominence as a highly active research area. LLMs are also applied in rescoring and fusion methods (Sections \ref{method-fusion} and \ref{method-rescoring}), but are consolidated in the same sections because their exploration in those contexts has remained comparatively limited.

The first line of work explores text-only prompting. Min and Wang~\cite{min2023exploring}  prompt an LLM with instructions, ten examples, and the 1-best ASR hypothesis to perform transcription correction, but this approach actually increases WER. Even when generating multiple corrected candidates and selecting the one with the lowest WER, the final performance still degrades compared to the original ASR output. After that, Yang et al.~\cite{yang2023generative} propose an LLM-based ASR correction method that improves n-best hypotheses through Task-Activating Prompting (TAP), a multi-round prompt in which the model is guided through a sequence of question-and-answer steps: first clarifying the task, then providing an example, and finally presenting the n-best hypotheses for correction. They compare TAP against zero-shot domain-hint prompting, where the LLM receives only a brief domain cue, and zero-shot chain-of-thought prompting \cite{wei2022chain}, which guides the model to reason step-by-step before correcting. They also evaluate few-shot and one-shot prompting, which provide example demonstrations. Across experiments, TAP consistently outperforms all these alternatives and even exceeds the oracle n-best WER, demonstrating its ability to correct errors beyond simple hypothesis selection. Additionally, they show that using parameter-efficient adapters, such as LoRA \cite{hu2022lora} and residual adapters \cite{houlsby2019parameter}, on open models with TAP-based input can further reduce WER, even surpassing full fine-tuning of the entire model.

Ma et al. \cite{ma2023can} explore the use of GPT-3.5 to post-correct ASR's n-best hypothesis lists via three strategies: unconstrained correction, where the model rewrites the transcription freely; selective correction, where ChatGPT picks the best hypothesis from all given hypotheses at once (this strategy is categorized into the rescoring Section \ref{method-rescoring}); and closest mapping, where the hypothesis nearest to the unconstrained output is chosen based on edit distance. Their findings demonstrate that this approach yields lower WER, even in out-of-domain scenarios, and that the improvements for transducer-based ASR can surpass those of a fine-tuned T5 using the n-best \cite{ma2023nbest}. Building on their earlier work, the authors in \cite{ma2025asr} apply the same three core strategies using both GPT-3.5 and the more capable GPT-4, while also evaluating these strategies with a supervised T5 model. They extend the hypothesis representation from n-best lists to full word lattices and introduce lattice-constrained decoding with the T5 model, which removes the limitation of a fixed number of hypotheses and allows exploration of a much richer hypothesis space. Furthermore, they demonstrate that merging hypotheses from two distinct ASR systems (i.e., a transducer-based model and Whisper \cite{radford2023robust}) before correction yields additional gains, effectively turning the LLM or T5 corrector into an ensembling mechanism across different ASR backends. Finally, they incorporate a simple reference-identification quiz to assess potential test-set contamination \cite{sainz2023nlp} in the pre-training data of the LLMs , reporting minimal contamination in GPT-3.5 and some level of contamination in GPT-4.

In the paper by Naderi et al. \cite{naderi2024towards}, the authors use an LLM to post-process ASR transcripts by applying three filtering strategies on ASR confidence: average sentence-level confidence below a threshold; lowest word-level confidence below a threshold; and prompting the LLM to correct only specific words with low token confidence. In contrast to approaches relying solely on native ASR confidence, Pu et al.~\cite{pu2023multi} introduce a multi-stage framework that derives uncertainty estimates from LM rescoring of n-best hypotheses. If the top hypothesis's normalized confidence exceeds a threshold, it is selected; otherwise, the hypothesis is refined by an LLM through rule-based prompting that constrains corrections to the words in the n-best list while preserving sentence structure and length. This yields improvement in WER reductions across various domains, even in zero-shot settings.

Unlike prior LLM-based ASR-correction methods that rely on a single prompt, Sachdev et al. \cite{sachdev2024evolutionary} design five initial prompts to correct the top-five ASR hypotheses and then refine them using EvoPrompt \cite{tong2025evoprompt}, an evolutionary prompt-optimization method. EvoPrompt treats prompts as a population and iteratively improves them through selection, crossover, and mutation, generating new candidates and retaining those that perform best. Their results show that this iterative, population-based evolution of prompts consistently improves error-correction performance.

Rather than correcting short utterance-level transcripts, Tang et al.~\cite{tang2025full} target error correction on full-text ASR outputs from long recordings. After constructing the dataset, they test two prompting strategies: one asking the LLM for full-text correction directly, and another asking it to output detected errors and corrected forms in JSON, applied either to the whole text or to shorter multi-sentence segments. The JSON-based approach performs better, producing fewer hallucinations because the structured format gives the model more control and constraint during correction. In a follow-up work, Tang et al.~\cite{tang2025chain} propose the Chain of Correction (CoC) method, which segments the full-text ASR output and iteratively feeds each segment to the LLM together with the entire transcript and all previously corrected segments. This step-by-step procedure maintains global context while reducing error propagation, resulting in improved segment-level correction accuracy on very long transcripts.

Beyond text-only instruction-based techniques, several studies incorporate auxiliary information or fine-tune LLMs to improve their effectiveness. Approaches leveraging phonetic information have shown particular promise: Udagawa et al.~\cite{udagawa2024robust} employed a conservative data-filtering procedure to exclude noisy ASR hypothesis–reference pairs (see Section~\ref{datasets-filtering}) and subsequently fine-tuned a Japanese LLM on transcripts augmented with phoneme sequences. Building on this phonetic insight, Li et al.~\cite{li2024pinyin} proposed Pinyin-Enhanced Generative Error Correction (PY-GEC), which inputs both ASR transcripts and their Pinyin representations into an LLM to address pronunciation-related errors in Chinese. Trained on synthetic ASR errors with a multitask objective that learns the conversion between text and Pinyin, PY-GEC relies only on the 1-best hypothesis and its Pinyin at inference time, outperforming text-only methods on homophone and phonetic-confusion benchmarks. Expanding the concept of leveraging auxiliary information beyond phonetics, Li et al.~\cite{li2024investigating} use information from other languages to improve error correction implicitly by fine-tuning a multilingual LLM on the 1-best hypothesis rather than relying solely on n-best lists from a single language. Their results demonstrate improved multilingual ASR performance and enable transfer of error-correction capability across languages that share writing systems, showing that error correction with only the 1-best hypothesis is both viable and beneficial for high- and low-resource languages. In contrast, Ghosh et al.~\cite{Ghosh2024Jun} take a completely different approach by conditioning an LLM on lip-motion visual features to perform generative error correction of n-best ASR hypotheses; they also fine-tuned the LLM with parameter-efficient multi-modal adapters.

Acoustic information has recently emerged as an effective auxiliary signal for enhancing generative error correction. Radhakrishnan et al. \cite{radhakrishnan2023whispering} propose Whispering LLaMA, which freezes the LLaMA \cite{touvron2023llama} weights and fine-tunes two adapters: one dedicated to self-attention for the text modality and another for fusing audio features extracted from the Whisper encoder. These acoustic features are injected through keys and values initialized from Whisper’s cross-attention layers and projected to match LLaMA’s dimensionality. To evaluate acoustic fusion in broader settings, Chen et al. \cite{chen2024s} explore three strategies for integrating acoustic information into LLaMA when correcting n-best hypotheses generated by an ASR system. The early-fusion strategy concatenates speech units derived from HuBERT~\cite{hsu2021hubert} or Wav2Vec 2.0~\cite{baevski2020wav2vec} with the word embeddings of each hypothesis before feeding them into the LLM. The mid-fusion strategy incorporates acoustic encodings by inserting adapters directly into the LLM, as demonstrated by the Whispering LLaMA architecture. However, both early and mid fusion are susceptible to modality laziness, where the model over-relies on a single modality and fails to achieve genuine multimodal gains. To address this issue, the authors introduce a late-fusion method that treats ASR logits as acoustic evidence and LLM logits as textual evidence. Although related to shallow fusion (see Section~\ref{method-shallow-fusion}), their approach differs in that the LLM processes the n-best hypotheses independently of the ASR. Because late fusion occurs at the decision stage, the method mitigates the modality-laziness problems inherent to earlier integration strategies. They further propose Uncertainty-Aware Dynamic Fusion (UADF), which first calibrates both ASR and LLM logits to counteract model overconfidence and then dynamically assigns fusion weights according to the LLM’s uncertainty. When the LLM is confident, its logits receive greater weight, and when its uncertainty increases, more weight is allocated to the ASR.

Rather than fusing acoustic information into a text-based LLM, Hu et al.~\cite{hu2024listen} employ a multimodal LLM, SpeechGPT~\cite{zhang2023speechgpt}, which is designed to process speech units and textual hypotheses jointly without requiring explicit cross-modal fusion. Their approach provides SpeechGPT with both speech units and n-best hypotheses and instructs it to reason over the two modalities simultaneously. Because the n-best hypotheses produced by ASR systems typically differ in only a small number of token positions, sending all hypotheses directly to the LLM would introduce substantial redundancy. To address this, the authors reformulate the n-best list as a cloze-style prompt. They align the hypotheses by identifying divergent positions, replacing each such position with a blank, and providing the corresponding set of candidate options for that blank, including a \texttt{<NULL>} token to account for hypotheses that do not contain a word at that position. However, cloze prompts introduce bias toward particular option IDs, which can distort the model’s selection behavior. Instead of performing computationally expensive inference over permutations of options, the authors mitigate this bias through logits calibration, disentangling option-ID biases from genuine model preferences using a validation set. Once the model selects an option for each blank, the resulting filled-in transcript may still contain errors, especially when all n-best hypotheses share the same incorrect word. To address these residual errors, the authors propose an additional correction stage in which the cloze-generated transcript is refined by SpeechGPT with the aid of the accompanying speech units. SpeechGPT is fine-tuned for both the cloze selection stage and the subsequent correction stage. With this two-step procedure, their method achieves WER performance that surpasses even the n-best oracle, demonstrating the effectiveness of leveraging a multimodal LLM without explicit acoustic–text fusion mechanisms.

\subsubsection{RAG-Integrated}
\label{sub:rag-integrated}
Correction methods require the integration of contextual information to enhance performance. A recent paradigm, initially developed for chatbots, utilizes retrieval mechanisms to incorporate customized information, resulting in more specific and contextually appropriate responses. Similarly, in ASR error correction, retrieval-augmented generation (RAG) techniques retrieve relevant knowledge based on candidate outputs, grounding the correction process in valid examples. However, the main challenge lies in determining what knowledge to include and how to incorporate it, for which different studies propose various solutions. Below, we summarize representative methods in this category.

% text based retrieval
To extend understanding of LLMs for ASR refinement, which was primarily evaluated in \cite{ma2023can}, \textit{GEC-RAG}~\cite{robatian2025gecrag} addresses ASR errors by first generating transcriptions from a black-box ASR system, then retrieving lexically similar examples from a knowledge base of ASR hypothesis--reference pairs using TF--IDF and cosine similarity. These text-based retrieved examples, which capture typical error patterns similar to those in the transcript, are incorporated into in-context learning prompts for an LLM (GPT-4o), exposing the model to error examples that improve its understanding of the task at hand.
Ghosh et al.~\cite{ghosh2024failing} introduced \textit{DARAG}, which augments the LLM input with both n-best hypotheses and a set of relevant retrieved entities. To this end, the authors first construct an index of named entities. For each refinement input, they compute SentenceBERT~\cite{reimers2019sentence} embeddings of the initial ASR transcript and retrieve the most similar entities from the index using cosine similarity. These retrieved entities, together with the n-best hypotheses, are then provided to an LLM that has been fine-tuned on pairs of real and synthetic ASR hypotheses with errors and their corresponding corrections (see Section~\ref{dataset-tts-asr}), enabling it to generate an improved transcript. This retrieval-augmented design helps the model correct entity-related errors while improving the overall quality of the ASR transcription.

% audio retrieval
Li et al.~\cite{li2024highprecision} proposed an approach that uses audio instead of text for retrieval. The intuition is that, in noisy domains, the ASR output may not be a good representation of the input utterance. Their method generates embeddings directly from the utterance, using an embedding network trained to align audio and text representations. At inference, the audio-derived embedding is used to query a candidate correction database, and retrieved candidates are scored by their embedding distance to the input audio. This strategy avoids the propagation of errors from inaccurate ASR hypotheses and achieves higher precision in correction.

% text and audio retrieval
Xiao et al.~\cite{xiao2025contextual} extended this idea by retrieving contextual information using both a text-based retriever (conditioned on the first-pass ASR hypotheses and powered by SentenceBERT embeddings) and an audio-based retriever (which employs a speech encoder, a text-to-audio encoder, and a learned joint module for audio-guided retrieval). The retrieved audio and text embeddings are then used to fine-tune an LLM, which is conditioned on the ASR hypotheses to generate a context-aware corrected transcript.

\subsection{Distillation}
\label{method-distillation}
Distillation in the context of LMs and ASR training is a technique where a complex, pretrained LM (the teacher) transfers its knowledge to a simpler ASR model (the student) during the training phase. The LM, trained on large-scale text corpora, injects linguistic knowledge into the ASR model to improve its performance, particularly in scenarios with limited paired speech-text data. This process enhances ASR accuracy by leveraging the LM’s deep syntactic and semantic understanding, without increasing computational cost at the inference time.

In~\cite{bai2019learn}, the authors integrate a RNNLM~\cite{mikolov2010recurrent}, which operates in a left-to-right manner, to enhance a Seq2Seq ASR model through distillation. The RNNLM generates soft labels in the form of token probability distributions, which serve as auxiliary supervision for the student ASR model. The loss function combines two components: a hard loss, defined as the cross-entropy between ASR outputs and ground truth transcriptions, and a soft loss that encourages the student to mimic the teacher’s distribution. The combined objective is expressed as:
\begin{equation}
\mathcal{L}_{\mathrm{KD}} = (1 - \lambda)\,\mathcal{L}_{\mathrm{hard}} + \lambda\,\mathcal{L}_{\mathrm{soft}},
\label{eq:kd_loss}
\end{equation}
where the soft loss is computed as:
\begin{equation}
\mathcal{L}_{\mathrm{soft}} = \sum_i p^{\mathrm{teacher}}_i \,\log \frac{1}{p^{\mathrm{student}}_i}
.
\label{eq:soft_loss}
\end{equation}
Since the teacher’s parameters are fixed during distillation, its output distribution \(p^{\mathrm{teacher}}_i\) is constant and can be dropped from the numerator of the logarithm in the KL-divergence, yielding this simplified form in equation \eqref{eq:soft_loss}. Teacher distribution is softened via temperature-scaled softmax:
\begin{equation}
p^{\mathrm{teacher}}_i = \frac{\exp(z_i / T)}{\sum_j \exp(z_j / T)},
\label{eq:teacher_softmax}
\end{equation}
where \(z_i\) denotes the pre-softmax activation for token \(i\).

Going beyond a unidirectional language model, Bai et al.~\cite{bai2019integrating} introduce a bidirectional language model, the Causal clOze completeR (COR), for distillation into a Seq2Seq ASR system. In COR, the outputs of left-to-right and right-to-left Transformer block stacks are concatenated and then fed into a subsequent fusion Transformer block. However, because COR combines the two directions only at a shallow level, Futami et al.~\cite{futami2020distilling} instead employ BERT to obtain a deeply bidirectional language model as the teacher. Their method further strengthens the teacher signal by providing both preceding and succeeding utterances as input to BERT, enabling it to generate more contextually informed soft targets for distillation.

Bai et al. \cite{bai2021fast} propose distilling BERT into a NAR ASR model, Listen Attentively and Spell Once (LASO) \cite{bai2020listen}. LASO includes a Position Dependent Summarizer (PDS) whose fixed-length positional queries attend to the high-level encoder representations and yield a fixed-length sequence for the non-autoregressive decoder. Because the output sequence must have a preset length, the tokenizer includes an \texttt{<eos>} token that appears at the tail of the output. To transfer linguistic knowledge from text to speech, distillation is performed at the representation level by minimizing mean squared error (MSE) between the LASO decoder outputs and the corresponding BERT representations rather than matching softmax probability distributions. 
Futami et al. \cite{futami2022distilling} distill BERT into CTC based ASR to enable the encoder to absorb bidirectional language model knowledge while preserving CTC’s fast decoding. Because BERT produces token-level outputs whereas a CTC model emits frame-level outputs, they compute a forced alignment by selecting the CTC path that maximizes the product of the forward and backward variables~\cite{graves2006connectionist}. This alignment is then used to associate contiguous frame segments with each token. To distill BERT knowledge, they define a loss in which, for each token, they sum the KL divergences between the BERT token distribution and the CTC model’s frame distributions over the frames assigned to that token.

Deng et al. \cite{deng2022improving} propose using Continuous Integrate-and-Fire (CIF)~\cite{dong2020cif} to align BERT representations with encoder representations of a CTC style model. In the CIF method, a small network predicts per-frame weights that are normalized so their sum equals the number of tokens in the transcription; when the running sum of weights exceeds one, a “fire” occurs and the weighted sum of the accumulated consecutive frames is emitted as the speech-derived representation for that token. This produces one speech vector per token, which can be matched to the corresponding BERT embedding using a cosine or MSE loss. The authors also present an alternative method for distilling BERT: given prior knowledge of the target token length, a positional encoding matrix is used to query the encoder outputs via cross-attention, and the cross-attention output for each positional query becomes the token representation to be matched with BERT. In addition to BERT distillation, they explore transferring GPT-2 knowledge by fusing GPT-2 token embeddings with speech features through cross-attention and optimizing a classification loss.

Hentschel et al. \cite{hentschel2024keep} propose attaching an auxiliary attention decoder during training that attends to intermediate encoder layers as well as to the final encoder layer of a CTC model. The auxiliary decoder produces token distributions from each encoder layer, and the authors apply KL divergence losses between those intermediate token distributions and the teacher BERT distributions. Because the auxiliary decoder is present only during training and removed at inference, the CTC model retains its original fast decoding while its encoder internal layers learn stronger text-aware representations through multi-layer distillation. 
Choi and Park \cite{choi2022distilll2s} distill a cross-lingual text model, InfoXLM \cite{chi2021infoxlm}, into a multilingual CTC-based ASR model, XLSR Wav2Vec 2.0 \cite{conneau2021unsupervised}. For the alignment needed for distillation, they apply a shrink mechanism \cite{chen2016phone} to remove blank frames, use nearest neighbor interpolation to match sequence lengths, and introduce a learnable linear projection layer so that speech encoder states and InfoXLM representations lie in a shared space suitable for representation-level MSE distillation. This approach effectively injects cross-lingual text knowledge into the CTC encoder and yields clear performance gains, especially for low-resource languages.

Kubo et al.\ \cite{kubo2022knowledge} propose a distillation framework that introduces an auxiliary regression network, used only during training, to map ASR internal states into the BERT token--embedding space for both Seq2Seq and neural transducer models.
For Seq2Seq models, for each token $i$, the regression network $R$ takes the decoder state conditioned on the target prefix, while the target is the corresponding BERT embedding $e_i$. The embedding distillation loss is
\begin{equation}
L^{\text{Emb}}_{\text{Seq2Seq}}
=
\sum_{i} d\!\left(R\!\left(\phi(X, y_{1:i-1})\right),\, e_i\right),
\end{equation}
where $d(\cdot,\cdot)$ is a distance function such as an $\ell_p$ norm. For transducer models, they define an expected regression loss that marginalizes over all alignments:
\begin{equation}
L^{\text{Emb}}_{\text{Trans}}
=
\sum_{i}\sum_{t}
q_i(t \mid X, y)\,
d\!\left(R(\phi_t, \psi_i),\, e_i\right),
\end{equation}
where $\phi_t$ is the acoustic feature at frame $t$, $\psi_i$ is the prediction-network feature for token $i$, and $q_i(t \mid X, y)$ is the alignment probability from the transducer forward--backward marginals.

Because evaluating the regression network for every frame--token pair is computationally expensive, they propose a practical approximation: first compute the expected acoustic input for token $i$,
\begin{equation}
\bar{\phi}_i
=
\sum_{t} q_i(t \mid X, y)\, \phi_t,
\end{equation}
and then apply a single token-wise regression loss
\begin{equation}
\tilde{L}^{\text{Emb}}_{\text{Trans}}
=
\sum_{i}
d\!\left(R(\bar{\phi}_i, \psi_i),\, e_i\right).
\end{equation}

In~\cite{lee2022knowledge}, the authors use multiple heads in the Acoustic Model (AM) of the hybrid ASR model: one for the hard loss and one for the soft loss, to separate the supervised learning and distillation objectives. This allows the distillation loss to be applied from LMs with different output units, and the heads are trained in a hierarchical multi-task fashion. This approach enables the use of LMs with different output units and improves robustness to the two hyperparameters \( T \) and \( \lambda \) in Equations \ref{eq:kd_loss} and \ref{eq:teacher_softmax}, due to the separation of the output heads. Similarly,~\cite{han2023knowledge} proposes a hierarchical distillation framework that transfers LM knowledge into both the acoustic and linguistic components of an E2E ASR model. To ensure token-level alignment between speech and text, they adopt a CIF-based ASR model~\cite{dong2020cif}, which produces one acoustic representation per token by integrating encoder outputs over time. At the acoustic level, contrastive, MSE, and cosine losses are used to align BERT token embeddings with CIF outputs, while at the linguistic level, an MSE loss aligns BERT representations with the ASR decoder's pre-softmax outputs.

\subsection{Training Adjustment}
\label{method-trainingadjustment}
% multi-task training, Minimum WER training & Smooth Labeling
Within the framework of non-intrusive ASR refinement, where the model architecture remains unchanged and no additional training data is introduced, optimizing the training process itself emerges as a key strategy for enhancing performance. This part surveys approaches that adjust the learning process to improve transcription accuracy, incorporate complementary training signals, or regulate model confidence for enhanced robustness and generalization.
\subsubsection{Internal Language Model Training}
\label{ta-ilmt}
In E2E ASR systems, the encoder and decoder are often regarded as the AM and LM, respectively. However, these relationships are implicit, and particularly the decoder may not behave as a true LM within the training domain, since it is conditioned on acoustic features during decoding. Therefore, improving the training scheme to mitigate this inconsistency has become an important direction of research.
% mention the method itself
Meng et al. in~\cite{meng2021internal} propose Internal Language Model Training (ILMT) that augments the standard E2E loss with an $ILM$ loss that encourages the acoustically‐conditioned decoder (for Attention-based Encoder-Decoder models) or the prediction and joint networks (for RNN‑T) to behave like a standalone LM.  In case of the RNN‑T, the ILM loss is defined as:
\begin{equation}
\mathcal{L}_{\mathrm{ILM}}(\theta)
= -\sum_{u=1}^{U}\log p\bigl(y_u \mid Y_{<u};\,\theta\bigr),
\end{equation}
thus the combined ILMT loss becomes,
\begin{equation}
\mathcal{L}_{\mathrm{ILMT}}(\theta_{\mathrm{RNN\text{-}T}})
= \mathcal{L}_{\mathrm{RNN\text{-}T}}(\theta_{\mathrm{RNN\text{-}T}})
\;+\;\alpha\,\mathcal{L}_{\mathrm{ILM}}(\theta),
\end{equation}
where \(Y_{<u}=Y_{0:u-1}\), \(\theta=(\theta_{\mathrm{pred}},\theta_{\mathrm{joint}})\), and $\alpha$ weights the ILM term.  For Attention-based Encoder-Decoder (AED) model, the ILM loss is
\begin{equation}
\mathcal{L}_{\mathrm{ILM}}(\theta_{\mathrm{dec}})
=  \sum_{u=1}^{U+1} 
\log p\bigl(y_u \mid Y_{<u};\,\theta_{\mathrm{dec}}\bigr),
\end{equation}
% - \sum_{Y\in\mathcal{D}}
and thus the ILMT loss is
\begin{equation}
\mathcal{L}_{\mathrm{ILMT}}(\theta_{\mathrm{AED}})
= \mathcal{L}_{\mathrm{AED}}(\theta_{\mathrm{AED}})
\;+\;\alpha\,\mathcal{L}_{\mathrm{ILM}}(\theta_{\mathrm{dec}}).
\end{equation}
By jointly minimizing \(\mathcal{L}_{\mathrm{E2E}} + \alpha\,\mathcal{L}_{\mathrm{ILM}}\), where \(\mathcal{L}_{\mathrm{E2E}}\) denotes \(\mathcal{L}_{\mathrm{AED}}\) for AED models and \(\mathcal{L}_{\mathrm{RNN\text{-}T}}\) for RNN-T models, ILMT updates only the decoder parameters (or the prediction and joint networks for RNN-T) for the LM term, while the rest of the end-to-end model is trained to optimize ASR accuracy.

\subsubsection{Minimum Word Error (MWE) Training}
\label{sec:mwe}
% mention the method itself and cite papers that use it
In another approach, the objective is to directly optimize error metrics. This process is typically applied as a fine-tuning stage following the main training procedure. 

Hori et al.~\cite{hori2016minimum} proposed a training strategy that adjusts the model parameters to minimize the expected number of word errors by computing a weighted sum of edit distances between the reference transcript and all hypothesized outputs, where the weights correspond to the posterior probabilities of the model. Specifically, for each utterance \(k\), given an \(N\)-best list \(\{W_{k,n}\}_{n=1}^N\) of candidate transcriptions based on acoustic observations \(O_k\), the MWE risk is defined as:
\begin{equation}    
    \mathcal{L}(\Lambda) = \sum_{k=1}^K \sum_{n=1}^N E\left(W_k^{(R)}, W_{k,n}\right) p_\Lambda\left(W_{k,n} \mid O_k\right),
\end{equation}
where \(E\left(W_k^{(R)}, W_{k,n}\right)\) represents the number of word errors (insertions, deletions, substitutions) between the reference sequence \(W_k^{(R)}\) and the n-th hypothesis \(W_{k,n}\), and \(p_\Lambda\left(W_{k,n} \mid O_k\right)\) is the posterior probability of each hypothesis according to the model parameters \(\Lambda\). 

Prabhavalkar et al. \cite{prabhavalkar2018minimum} developed MWE training for attention-based sequence-to-sequence models. They propose minimizing the expected number of errors on the training data:
\begin{equation}
    \mathcal{L}_{\mathrm{werr}}(x,y^*) \;=\; \mathbb{E}_{p(y\mid x)}\big[\mathcal{W}(y,y^*)\big]
    \;=\; \sum_{y} p(y\mid x)\,\mathcal{W}(y,y^*),
\end{equation}
where \(\mathcal{W}(y,y^*)\) is the number of errors in a hypothesis \(y\) relative to the reference \(y^*\). Computing the sum over all possible labels is intractable, so they approximate the expectation using two methods.
First, empirical averaging over \(M\) samples drawn from the model:
\begin{equation}
    \mathcal{L}_{\mathrm{werr}}(x,y^*) \approx 
    \mathcal{L}_{\mathrm{werr}}^{\mathrm{Sample}}(x,y^*)
    \;=\; \frac{1}{M}\sum_{i=1}^{M} \mathcal{W}(y_i,y^*), \quad y_i \sim p(y\mid x).
\end{equation}
Second, restricting the sum to an n-best (beam) list and using a normalized posterior; a baseline \(\hat{W}\), which is the average number of word errors over the n-best hypotheses, is subtracted to reduce variance:
\begin{equation}
    \mathcal{L}_{\mathrm{werr}}(x,y^*) \approx 
    \mathcal{L}_{\mathrm{werr}}^{\mathrm{N\text{-}best}}(x,y^*)
    = \sum_{y_i \in \mathrm{Beam}(x,N)} 
      \frac{p(y_i\mid x)}{\sum_{y_j \in \mathrm{Beam}(x,N)} p(y_j\mid x)}
      \big[\mathcal{W}(y_i,y^*) - \hat{W}\big].
\end{equation}
The authors report that optimization using the n-best approximation outperforms the sampling estimator and yields consistent improvements on a held-out dataset.
Moreover, the final training objective interpolates the expected-WER loss \(\mathcal{L}_{\mathrm{werr}}(x,y^*)\) with the cross-entropy loss \(\mathcal{L}_{\mathrm{CE}}\), demonstrating the importance of including the CE term. The interpolated objective is given in Equation~\ref{eq:mwe-interpolate}.

\begin{equation}
\label{eq:mwe-interpolate}
    \mathcal{L} = \sum_{(x, y^*)} \mathcal{L}_{\text{werr}}(x, y^*) + \lambda \mathcal{L}_{\text{CE}}
\end{equation}

Therefore, by directly optimizing the parameters of the ASR system, MWE training targets the reduction of transcription errors and leads to statistically significant improvements in WER compared to conventional cross-entropy training.

\subsubsection{Label Smoothing (LS)}
% mention the method itself
An effective remedy for the tendency of sequence-to-sequence ASR models to become over-confident \cite{kuhn2025evaluating} is \emph{labeling smoothing}, in which the one-hot ground-truth distribution is relaxed by assigning a fraction $\beta$ of the probability mass to the correct token and redistributing the remaining $1 - \beta$ across plausible alternatives. Beyond \emph{uniform} proposed in~\cite{szegedy2016rethinking} and \emph{unigram} in~\cite{pereyra2017regularizing}, smoothing in \cite{chorowski2016towards} introduces \emph{neighborhood smoothing}, wherein the residual probability is concentrated on tokens within $\pm1$ and $\pm2$ positions in the reference transcript, thereby reflecting the temporal structure of speech.  By discouraging the model from collapsing its probability mass onto a single token, neighborhood smoothing both regularizes learning, yielding nearly a 3\% reduction in character error under greedy decoding, and increases the entropy of the output distribution.  A small grid search over $\beta$ settings demonstrates robust performance across a broad range of parameter choices, underscoring the practical value of smooth labeling in ASR decoding.

\subsection{Comparative Discussion}
The five classes of non‑intrusive refinement span a wide spectrum of trade‑offs, and their practical value depends strongly on the deployment context. We first define the comparison axes, then provide a method‑by‑method synthesis of strengths, frailties, and suitable scenarios, and finally discuss cross‑cutting themes and offer a concise selection guide.

\subsubsection{Dimensions of Comparison}
We evaluate each category along the following dimensions (some of which are mentioned in Table \ref{tab:method-comparison}): (i) ASR retraining required, (ii) inference latency and memory, (iii) potential error reduction, (iv) robustness and over‑correction risk, (v) external data dependence (need for text‑only corpora or paired erroneous-correct transcription data), and (vi) integration complexity. The discussion below adds qualitative nuance that cannot be captured in a single table (Table \ref{tab:method-comparison}).

\subsubsection{Method‑by‑Method Comparative Profiles}
\paragraph{\textbf{Fusion}}
Shallow fusion is the simplest integration path: it requires no ASR retraining and can exploit any pre‑trained LM as a plug‑and‑play source of linguistic bias during beam search. Typical relative WER reductions is moderate \cite{toshniwal2018comparison}, but the improvement is constrained by the mismatch between the external LM and the internal LM prior of the E2E model. Cross‑domain performance often suffers unless explicit bias‑correction techniques (DRM, ILME) are applied, which add hyperparameters and implementation overhead \cite{McDermott2019,Meng2021-estimation}. Deep and cold fusion offer tighter coupling, yet they necessitate partial or full retraining of the ASR decoder and have produced diminishing gains in recent studies \cite{toshniwal2018comparison}. Memory and latency are moderate for standard LMs but become prohibitive when LLMs are used inside the decoding loop. Overall, shallow fusion is well suited when the ASR cannot be altered, a small yet reliable improvement is desired, and the target domain matches the external LM well.

\paragraph{\textbf{Rescoring}} Rescoring enriches the ASR’s initial hypotheses with a stronger LM in a separate pass. First‑pass lattice rescoring can improve hypothesis quality before beam pruning, but it requires highly optimised graph‑search algorithms (push-forward, arc-beam, and clustering) to remain tractable \cite{auli2013joint,kumar2017lattice,liu2016two}. Second‑pass n‑best rescoring is far lighter and can leverage large transformers or even LLMs \cite{ma2023can,shivakumar2025speech}, yet its performance is capped by the n‑best oracle (if the correct transcript is missing, no re‑ranking can recover it). Modern discriminative rescoring models (e.g., RescoreBERT, PBERT) typically demand fine‑tuning on ASR hypothesis lists but consistently outperform zero‑shot scoring with pre‑trained LMs \cite{chiu2021innovative,xu2022rescorebert}. Rescoring is the method of choice when a diverse n‑best list can be obtained and a light second‑pass model can be trained, or when an LLM is to be exploited.

\paragraph{\textbf{Correction}} Correction frameworks can generate entirely new word sequences, thereby recovering words absent from the initial hypotheses. This capability yields the largest absolute WER reductions potential, especially when acoustic, phonetic, or retrieved contextual features supplement the textual input \cite{peyghan2025cmc,fan2023boosting,robatian2025gecrag}. However, the power to rewrite carries a constant risk of over‑correction—semantically plausible but acoustically unsupported substitutions—and, in the case of LLM‑based correctors, factual hallucinations \cite{min2023exploring}. AR decoder models are potent but slow; NAR variants dramatically reduce latency at the cost of some output fidelity. Most correction methods require paired erroneous‑correct training examples, and the quality of edit‑distance alignments used to create such pairs strongly influences downstream performance (particularly when a detection module is utilized to determine the erroneous tokens before the correction process) \cite{he2023ed,he2025pmf}. Data scarcity and domain mismatch can lead to brittle models that inadvertently alter correct words. Correction is most appropriate when the ASR’s raw error rate is high, when post‑processing latency is tolerable, and when sufficient in‑domain error data can be curated.

\paragraph{\textbf{Distillation}} Knowledge distillation transfers linguistic knowledge from a large and rich teacher LM into the ASR model’s encoder or decoder during training. Its primary advantage is that the refined ASR imposes no additional inference cost. This is especially valuable for low‑resource languages or on‑device models where inference‑time, memory and latency are severely restricted. The main drawbacks are (i) the need for full retraining of the ASR, which is impossible for black‑box services, (ii) the engineering challenge of aligning frame‑level acoustic representations with token‑level teacher embeddings \cite{futami2022distilling,chen2016phone}, and (iii) the distilled knowledge is static (it cannot be updated without re‑distillation). 
Distillation is ideal when retraining is permitted, when inference‑time resources are extremely tight, and when a strong text‑only LM is available but cannot be used during decoding.

\paragraph{\textbf{Training Adjustment}} Methods that refine the training objective (ILMT, MWE training, label smoothing) are genuinely lightweight: they introduce zero inference overhead and zero additional parameters. ILMT encourages the decoder to behave as a well‑calibrated LM, which not only improves the ASR itself but also makes subsequent shallow fusion more effective \cite{meng2021internal}. MWE training directly optimises the metric of interest and often yields a small relative WER gain after standard CE training \cite{hori2016minimum}. Label smoothing mitigates over‑confidence (a more diverse n-best list that increase the potential of rescoring method) and improves generalisation \cite{szegedy2016rethinking,pereyra2017regularizing,chorowski2016towards}. The trade‑off is that retraining is required, and the gains, while reliable, are typically smaller than those of methods that introduce a much stronger external LM. Moreover, ILMT and MWE add extra loss terms whose weighting must be carefully tuned to avoid competing with the primary ASR objective. These adjustments are best viewed as complementary fine‑tuning steps that can be applied before other non‑intrusive methods to prepare a stronger base model.

\subsubsection{Practical Guides}
Based on the above analysis, we offer the following heuristics to navigate the design space.

\begin{itemize}
    \item \textbf{If the ASR is a black‑box service (no access to internal scores or decoding):} Methods that require integration into the decoder (shallow fusion, deep fusion, cold fusion) are not applicable. The practical options are second‑pass rescoring (\ref{method-rescoring}) (if the service returns an n‑best list with scores) and correction (\ref{method-correction}) (which can operate directly on the final transcript or the n-best list). Zero‑shot and few‑shot LLM-based correction and rescoring techniques are particularly attractive here because they require no additional training data and can work with just the 1‑best or n‑best output \cite{robatian2025gecrag,ma2023can,shivakumar2025speech,yang2023generative,li2024highprecision}. Plus, if the black‑box ASR provides only a 1‑best transcript, correction is the only applicable path.
    
    \item \textbf{If the ASR model is accessible (white‑box) but retraining is not applicable:} Shallow fusion (\ref{method-fusion}) and first-pass rescoring become available in addition to correction and second-pass rescoring. Shallow fusion offers a small but reliable gain without altering the ASR, while rescoring can exploit a stronger LM on the lattice or the n‑best list.
    
    \item \textbf{If latency is paramount:} Prefer training adjustments (\ref{method-trainingadjustment}), distillation (\ref{method-distillation}), or lightweight correction methods (encoder-based, and NAR techniques in \nameref{sub:nlm-correction}). Avoid large AR decoders and LLM scorers inside the decoding loop; consider two‑pass rescoring with a very compact second‑pass model.
    
    \item \textbf{When high accuracy on rare or domain‑specific terminology is required:} Correction methods using RAG system (\nameref{sub:rag-integrated}) or the phrase‑capturing adaptation techniques (\ref{sec:adaptation}) are most effective. These can actively retrieve and inject the correct terminology into the output.
    
    \item \textbf{For languages with limited paired speech data but abundant text corpora:} Distillation (\ref{method-distillation}) and shallow fusion (\ref{method-fusion}) offer the best return, as they leverage text‑only resources to compensate for the lack of audio‑transcript pairs.
    
    \item \textbf{To build a robust ASR system where the ASR model is available and retraining is feasible:} A multi‑step paradigm is recommended. Start by applying ILMT and MWE training to the base ASR to improve its internal linguistic calibration, and use label smoothing to cope with the model’s over‑confidence (\ref{method-trainingadjustment}). If latency permits, shallow fusion (\ref{method-fusion}) can be added for further improvement. Then employ discriminative rescoring (\ref{method-rescoring}) on the n‑best list and reserve a heavier LM or LLM corrector (\ref{method-correction}) for cases where the rescoring confidence is low or where the n‑best list entropy indicates ambiguity. This layered strategy keeps average latency low while exploiting the full capacity of large LMs exactly where they are needed most.
    
\end{itemize}

%%%%%%%%%%%%%%%%%%%%%%%%%%%%%%%%%%%%%%%%%%%%%%%%%%%%
%     C O N T E X T U A L   B I A S I N G
%%%%%%%%%%%%%%%%%%%%%%%%%%%%%%%%%%%%%%%%%%%%%%%%%%%%

% Starting the Adaptation section
\section{Adaptation}
\label{sec:adaptation}
General ASR models often misrecognize specialized terminology and rare words in domain-specific contexts due to their low occurrence in general training corpora, which biases model predictions toward phonetically similar but more frequent tokens. Consequently, adaptation techniques become essential for optimizing ASR performance in rare-word and low-resource scenarios. Various non-intrusive strategies have been proposed to address this issue, which are mostly customized versions of the previously mentioned refinement techniques. Here, we categorize these methods into three main trends: (i) re-training with domain-specific synthesized paired or text-only data, (ii) incorporating predefined lists of domain-relevant terminologies, and (iii) leveraging the generalization capabilities of LLMs. This section examines these approaches, their effectiveness, and the challenges they face in improving ASR performance across diverse application domains.

% Subsection for Domain-Specific Training
\subsection{Domain-Specific Training}

This approach involves re-training or fine-tuning ASR models or external LMs with domain-specific synthesized paired or text-only data to improve performance in a target domain. A key challenge is the limited availability of domain-specific data, particularly for retraining ASR systems. For external LMs used in the correction tasks, the scarcity of domain-specific noisy-clean text pairs (generated by ASR) poses a similar issue, prompting innovative solutions to mitigate this constraint which will be discussed below.

Mai et al. \cite{mai2022unsupervised} address the data scarcity challenge by introducing an unsupervised method that employs an encoder-decoder correction network. This approach uses pseudo-labeled data, generated by transcribing audio with inferior ASR models (sub-optimal inputs) and a superior model (optimal targets), to train the correction network without ground truth labels, enhancing domain-specific performance.
Nanayakkara et al. in \cite{nanayakkara2022clinical} also focus on the medical domain, evaluating seq2seq encoder-decoder models for post-error correction. They highlight the poor performance of pretrained LMs in clinical contexts and demonstrate the need for fine-tuning with domain-specific text-only data. To overcome the lack of ASR transcribed data, they explore fine-tuning tasks like summarization, paraphrasing, and mask-filling, with mask-filling proving most effective due to its inherent similarity with error correction objectives. % Seq2Seq / Class-room domain
Recently, Jia et al., in Error Pattern Informed Correction (EPIC)~\cite{Jia2025EPIC}, fine-tuned a pretrained LM on classroom-specific synthetic data generated by an LLM based on real ASR error patterns. The LM was fine-tuned and, during inference, generated multiple correction candidates for each input. A learned similarity scoring mechanism was then applied to select the most contextually plausible correction among them.

% Subsection for Capturing Specific Phrases or Words
\subsection{Capturing Specific Phrases or Words}
\label{sec:capturing-specific-phrases-words}

The methods within this trend prioritize the accurate recognition of a predefined list of phrases or words, typically without the broader objective of enhancing overall ASR performance. These approaches are particularly valuable in domains involving critical terminology, although their effectiveness is constrained by the scope of the provided list. Nevertheless, some of the methods discussed below incorporate selection mechanisms to manage large corpora and improve scalability.
The structure of this section aligns with that of the main section in the paper (see Section \ref{sec:methods}). This is because the techniques presented here represent applications of the methods already described. In fact, the approaches in this section introduce algorithms for integrating a list of specific entities into the refinement process, ensuring that the refinement methods prioritize them.

In an early method, Sarma and Palmer \cite{sarma2004context} proposed an unsupervised method to detect and correct ASR errors on critical keywords with high precision. The approach first learns characteristic contextual patterns for each keyword by computing co-occurrence statistics from large ASR transcript corpora. During inference, it scans transcripts with a sliding window, identifying words whose surrounding context matches that of a target keyword and correcting them only when strong phonetic similarity is also present.

Anantaram et al.\ \cite{anantaram2018repairing} propose a multi-stage repair pipeline. 
First, a sliding \(n\)-gram window scans the transcript and attempts to match spans to a elements of a domain-specific list using exact string matches, or a combination of phonetic similarity  Levenshtein distance. Second, selected candidate matches from the previous step are re-evaluated in context and phonetically-close words are mapped to the best-fitting domain term. Third, any remaining segments are passed to an LSTM seq2seq model (trained on domain data) to predict improved replacements. Finally, a rule-based linguistic pass (part of speech heuristics or external tools such as the open source LanguageTool) fixes residual grammatical inconsistencies. 

Zhao et al.~\cite{zhao2019shallow} introduce a shallow-fusion-based method for ASR contextual biasing that assumes prior knowledge of a list of biasing words, such as user-specific contacts, or names. It creates a contextual LM as a Weighted FST by composing a word-level n-gram FST \(G\) with a speller FST \(S\) to map to sub-word units (graphemes or wordpieces). Biasing is performed at each decoding step before pruning to help rare biasing phrases survive early beam search elimination. The approach also incorporates mined prefixes (e.g., "call" or "play") concatenated with the FST to constrain activation and avoid anti-context (non-biasing phrases) degradation. Moreover, it penalizes partial matches through subtractive costs on failure arcs where the full biasing phrase is not completed. 

Huang et al.~\cite{huang2025neural} introduced another fusion-based approach for contextual biasing which employs an AR attention-based biasing decoder that processes acoustic features from the ASR encoder to compute the conditional probability of candidate specific phrases \(P(p_i\mid O)\), while enabling a filtering mechanism that discards the majority of unlikely biasing phrases not included in the target transcription (distractors) to significantly reduce computational costs during decoding. The method calculates discriminative per-token scores for each phrase in the predefined list, which are then integrated via shallow fusion to provide bonuses during ASR beam search, based on \cite{wang2024contextual}, ensuring plug-and-play compatibility with any ASR architecture. Evaluations demonstrate improvement on biasing phrases while maintaining or minimally affecting WER on unbiased general phrases.

% bert based
Kolehmainen et al.~\cite{kolehmainen2023personalization} suggest that general rescoring technique (i.e., RescoreBERT~\cite{xu2022rescorebert}) struggle to handle personalized content such as named entities (for example, contact names), and can often degrade performance on such specialized domains compared to the first-pass model. To address this, the authors propose three extensions: (i) \textit{prompting}, where natural-language prompts (e.g., ``as I need to contact [entity]'') are concatenated to hypotheses that match any personalized entity from a user's set \(\mathcal{D}\); (ii) \textit{gazetteers}, which tag matching tokens in hypotheses with binary indicators (1 for full matches to entities in \(\mathcal{D}\), 0 otherwise) and add trainable slot embeddings either early (to the input embeddings before the first transformer layer) or late (to the final layer); and (iii) a cross-attention-based encoder–decoder, in which matched entities are concatenated into a slot sequence, encoded separately, and attended to through cross-attention in a decoder layer. Among these methods, the gazetteer approach provides the best performance, achieving over \(10\%\) relative improvement on a personalized test set (which contains samples with personalized terms) while also improving results on a general test set. The prompting scheme also shows about a \(7\%\) relative improvement in WER over the first-pass model without any additional training, making it suitable when personalized data is scarce. The cross-attention method was unable to find any configuration that improved the personalized set without degrading performance on the general set.

Wang et al. \cite{wang2022towards} present a contextual spelling correction (CSC) framework based on neural encoder-decoders with two variants: an AR CSC (AR-CSC) and a NAR Fast CSC (FCSC). In AR-CSC, a text encoder encodes the original ASR transcript, a context encoder creates context embeddings from context phrases, and the decoder attends to the embeddings of both encoders and autoregressively decodes a context-aware corrected transcript. In FCSC, the same modules exist; however, the decoder provides two different outputs: a tag sequence that has the same length as the ASR transcript and determines whether a token in the transcript is a candidate for correction, and a similarity head that finds which context phrase to use based on the similarity between the context embeddings and the embeddings of the source transcript. Moreover, to handle the scalability of context phrases, the framework proposes two rankers, preference and relevance, as a selection mechanism to choose the most plausible list of phrases for each correction process.

He et al.~\cite{he2023ed} introduce ED-CEC, a two-branch framework that improves rare-word recognition by detecting and contextually correcting errors. A BERT-based classifier, resembling the detection module in \cite{yang2022asr}, identifies error positions in the source input. The input is represented as the sequence of initial transcript tokens with \(\langle u_i\rangle\) inserted between each pair of consecutive tokens to handle ASR deletions. An autoregressive decoder then decodes only these error positions, improving both the speed and efficiency of inference. The model augments the general decoder output with contextual embeddings for rare words via a context decoder. Finally, a weighted summation of the general and context decoder outputs is used to determine which rare words appear in the final corrected transcription.
Despite the advances achieved by this technique, relying on text-only processing may fall short when handling rare words that are phonetically similar but spelled differently (homophones). He et al.~\cite{he2025pmf} therefore extend their previous work by adding contextual phonetic embeddings to both the detection and correction modules, where phonetic representations are extracted from the source text. To handle the length mismatch between text and phonetic sequences, they introduce a Phoneme-augmented Multimodal Fusion (PMF) module, which leverages a cross-attention layer to learn alignment between the modalities and produce multimodal embeddings.

Antonova et al.~\cite{antonova2023spellmapper} develop SpellMapper, a non-autoregressive spellchecker that uses a pre-built mapping index, created via GIZA++ alignments on TTS-synthesized and ASR-decoded phrase pairs, to correct misspelled phrases. It first selects the correction candidates using character-level n-grams and then decides on candidates of replacement using a BERT-based classifier. This enables domain-customized corrections without modifying the ASR, though it requires an initial preprocessing step to create the mapping index.

A well-designed paradigm to include external knowledge into the refinement process is the use of RAG systems. Different methodologies have adopted RAG to improve ASR output on a list of specific phrases. 
Pusateri et al.~\cite{pusateri2025retrieval} proposed a RAG-based approach for correcting entities in ASR transcripts. The method begins by constructing a list of domain-specific entities. The ASR transcript is then analyzed to extract potential entity mentions, using several retrieval strategies. These candidate entities are embedded to jointly capture semantic and acoustic information and are compared against embeddings of the predefined entity list. To avoid overly long prompts, only the most relevant retrieved entities are retained, aligned with the ASR transcript, and finally provided to the LLM for correction.

Wang et al.~\cite{wang2024dancer} proposed DANCER, a method that detects potential entity mistakes in ASR transcripts and replaces them with corrected entities. To achieve this, they first compile a list of entities along with their descriptions. They then employ a classifier to identify characters in the transcript that are candidates for correction. Next, a phonetic-based retrieval system identifies correction candidates for each error-labeled span in the transcript. A context encoder processes the masked transcript (where the error span is masked), while an entity encoder handles the retrieved entities (including their descriptions). Based on mutual information and phonetic similarity, the candidates are re-ranked to determine the most appropriate entity for the masked span. Finally, an entity rejection module is applied to reject any candidate entities that do not match the phonetics of the corresponding span in any of the n-best hypotheses.
Luo et al.~\cite{luo2025generative} introduced a generative correction approach that leverages TTS synthesis to improve accuracy on domain-specific entities. By encoding TTS-generated entity audio using the ASR encoder and comparing these embeddings with those of input speech segments, the method enables speech-based retrieval to identify entity matches in the input. The top retrieved candidates, together with the source transcript, are then fed into a decoder model, which is conditioned on audio embeddings from the ASR encoder to generate correction candidates for each selected entity (including the option to produce an ``empty'' output when rejecting an initial retrieval). This effectively addresses cases where ASR errors exhibit substantial phonetic divergence from the correct entities, in which scenarios text-derived phonetic cues are insufficient~\cite{raghuvanshi2019entity,garg2020hierarchical}.

All previous RAG-based methods for entity correction may experience confusion when the number of retrieved candidates is high, necessitating techniques to filter noisy retrieved examples and retain the most relevant ones. To address this, \textit{DeRAGEC}~\cite{im2025deragec} proposes synthesizing rationales for training data that explain how each candidate entity contributes to correcting the ASR hypothesis. These rationales, along with the n-best hypotheses, target transcripts, candidate entities, their phonetic similarity scores, and one-sentence descriptions, augment the dataset for use as few-shot examples for in-context learning (ICL). During inference, for each input hypothesis, the phonetically retrieved named entity candidates are filtered using this ICL approach, providing better guidance to the LLM for the correction process by enhancing its understanding of each entity's impact and reducing confusion.
% neural network biasing
In another recent work, Zhou et al. \cite{zhou2025autodrafting} propose a human-in-the-loop approach (which is to some extent a RAG-based system while the external knowledge is directly added by human supervision) for generating police reports, ensuring accuracy by allowing users to enter or verify specific keywords. This method relies on manual input to capture critical terms, bypassing extensive model retraining.

% Subsection for LLM-Based Methods with Prompting
\subsection{LLM-Based Domain Generalization}

The main approach in this category leverages the generalization capability of LLMs to refine ASR outputs across different domains through diverse prompting schemes, often without the need for domain-specific training. This strategy exploits the extensive knowledge embedded in LLMs and adapts them to specific contexts via tailored prompting techniques.

Ebadi et al.~\cite{ebadi2024extracting} employed zero-shot and few-shot prompting with LLMs to enhance transcription accuracy in the medical domain, further evaluating the improvements in named entity recognition tasks to demonstrate the adaptability of LLMs for domain-specific refinement without retraining. Adedeji et al.~\cite{adedeji2024sound} applied Chain-of-Thought (CoT) prompting for ASR post-refinement in the same domain, achieving superior performance over zero-shot prompting. This method, extending their broader research in \cite{adedeji2025multicultural}, leverages structured reasoning to enhance correction accuracy across diverse application domains.

\section{Datasets}
\label{sec:datasets}

To train and evaluate models for ASR error correction, specialized appropriate datasets are crucially required. 
Approaches for dataset creation range from collecting real-world benchmarks derived from actual ASR system outputs and their corresponding corrected versions \cite{Chen2023Sep}. These datasets often reflect errors specific to certain domains, languages, accents, ASR systems, speaker variations, or acoustic conditions. Another common method is generating data synthetically using Text-to-Speech (TTS) systems followed by ASR processing to create paired noisy/clean text. Furthermore, pseudo-data is frequently constructed by introducing synthetic errors into large volumes of clean text, often leveraging LLMs or observed error patterns to mimic ASR outputs.
This section details the diverse methodologies employed to create and utilize such datasets, which are fundamental for advancing the field. Furthermore, strategies for filtering and refining these datasets to enhance the robustness and performance of ASR refinement models are discussed, ensuring that the training data accurately reflects the complexities and nuances of real-world ASR errors. 
Finally, Table~\ref{tab:datasets-tr} presents the text refinement datasets and ~\ref{datasets-tables} provides an extensive list of datasets originally designed for speech recognition, which serve as valuable resources for ASR refinement and error-correction studies.

\begin{table*}[t]
\centering
\caption{Datasets used for Text and Transcript Refinement. Language abbreviations in the table are: EN (English), ZH (Chinese), and ML (Multi-Lingual).}
\label{tab:datasets-tr}

\begingroup
\renewcommand{\arraystretch}{1.3}
\fontsize{9}{10}\selectfont

\begin{tabularx}{\textwidth}{p{2.3cm} c c p{2.6cm} c X c}
\toprule
\textbf{Dataset} &
\textbf{Year} &
\textbf{Lang.} &
\textbf{Primary Task} &
\textbf{\#Samples} &
\textbf{Notes} &
\textbf{Ref.} \\
\midrule

HyProdise (HP) & 2023 & EN & Text refinement & 334K &
General (mixed) domain &
\cite{chen2023hyporadise} \\

SIGHAN13 & 2013 & ZH & Text refinement & 1.7K &
Language learning (TOCFL), L2 learners &
\cite{wu2013chinese} \\

SIGHAN14 & 2014 & ZH & Text refinement & 4.5K &
Language learning (TOCFL), L2 learners &
\cite{yu2014overview} \\

SIGHAN15 & 2015 & ZH & Text refinement & 3.4K &
Language learning (TOCFL), L2 learners &
\cite{tseng2015introduction} \\

Wang271K & 2018 & ZH & Text refinement & 271K &
General domain (news-inclusive); OCR- and ASR-style errors &
\cite{wang2018hybrid} \\

LEMON & 2023 & ZH & Text refinement & 22K &
Human-written text; manual transcripts; multi-domain (medical-inclusive) &
\cite{wu2023rethinking} \\

ChFT & 2024 & ZH & Text refinement & 562K &
% Full-text Chinese ASR error-correction dataset built using a TTS + ASR pipeline; segmented from THUCNews articles (41,651 training articles; test sets: 5,592 homogeneous, 3,987 up-to-date, 5,590 hard); samples (segments) contain multiple sentences and are extracted from their corresponding full articles; covers news and broad domains, including punctuation restoration and inverse text normalization (ITN) errors &
Built using a TTS + ASR pipeline; covers news and broad domains, including punctuation restoration and inverse text normalization (ITN) errors &
\cite{tang2025full} \\

ASR-EC & 2024 & ZH & Text refinement & 1M &
General (mixed) domain &
\cite{wei2024asr} \\

CSCD-NS & 2024 & ZH & Text refinement & 40K &
Social media posts; native speakers &
\cite{hu2024cscd} \\

CoVoGER & 2025 & ML & Text refinement & 3.34M &
General (mixed) domain; n-best hypotheses &
~\cite{yang2025covoger} \\

\bottomrule
\end{tabularx}
\endgroup
\end{table*}

\subsection{Real-World Benchmark Datasets}

Real-world benchmark datasets for ASR refinement are predominantly constructed by leveraging pretrained ASR models or by training custom ASR models on public or private speech datasets to generate ASR ground-truth pairs $\text{\cite{hu2020deliberation, Ghosh2024Jun, Wang2018}}$. Most research in this area utilizes established ASR systems to transcribe speech corpora, creating paired datasets of ASR hypotheses and their corresponding ground-truth transcripts. These datasets are designed to capture authentic ASR errors that arise from diverse real-world conditions, such as background noise, variations among speakers, and acoustic distortions. To foster broader research adoption, some researchers publicly release these generated datasets. Furthermore, many approaches incorporate data augmentation techniques, including the addition of background noise, speed perturbation, and other acoustic modifications, to further introduce realistic ASR errors and enhance dataset diversity, making them suitable for evaluating model performance in practical scenarios.

Three significant examples of datasets created using industry-grade ASRs to generate (ASR,ground truth) pairs are HyPoradise ($\text{HP}$), ASR-EC, and CoVoGER. Chen et al. $\text{\cite{Chen2023Sep}}$ developed the $\text{HP}$ dataset by generating n-best hypotheses (top-5 from a beam search with a beam size of 60) from two state-of-the-art ASR models, WavLM and Whisper \cite{radford2023robust}. This process has been applied to numerous popular ASR datasets, including LibriSpeech \cite{7178964}, CHIME-4 \cite{Vincent2016Chime4}, WSJ \cite{Paul1992WSJ}, SwitchBoard \cite{godfrey1997switchboard}, Common Voice \cite{Ardila2019DecCV}, Tedlium-3 \cite{Hlubik2020Tedlium}, LRS2 \cite{Chung2016NovLRS2}, ATIS \cite{Hemphill1990ATIS}, and CORAAL \cite{coraal2021}, yielding over 334,000 pairs. In a similar vein, the ASR-EC benchmark \cite{wei2024asr}, specifically designed for Chinese ASR error correction, was constructed by collecting erroneous transcriptions from industry-grade ASR systems and pairing them with manually verified ground-truth transcripts. ASR-EC utilizes data from diverse Chinese speech corpora, such as AISHELL-1 \cite{bu2017aishell} and WenetSpeech \cite{Zhang2021OctWenetspeech}, to cover a wide spectrum of ASR errors.
Recently, Yang et al.~\cite{yang2025covoger} created CoVoGER, a multilingual multitask benchmark for ASR and speech-to-text translation (ST) GEC (denoted as GER in the original paper) with LLMs. They decoded Common Voice 20.0 and CoVoST-2~\cite{wang2021covost} datasets using three Whisper model sizes (for ASR) and two SeamlessM4T model sizes~\cite{barrault2023seamlessm4t} (for ST), adopting beam search, temperature sampling, and a mixture of both. They conducted this process for 15 languages in ASR (14 non-English languages plus English) and 28 language pairs in ST (bidirectional between the 14 non-English languages and English), and provided the 5-best lists paired with the target transcript to form the benchmark.

\subsection{TTS-ASR Synthesis Methods}
\label{dataset-tts-asr}
TTS-ASR synthesis methods offer a versatile and scalable framework for generating synthetic datasets to enhance ASR error correction, particularly excelling in context-specific scenarios and full-text applications. This approach begins with converting clean text into speech using TTS systems, which can be fine-tuned to incorporate diverse variables such as speaker identity, accent, and background noise. The resulting audio is then transcribed back into text by an ASR model, yielding noisy hypotheses that replicate real-world transcription errors. This flexibility enables the creation of tailored datasets for training robust refinement models, accommodating challenges like multi-speaker environments or domain-specific nuances. By systematically manipulating TTS parameters, researchers can simulate a broad spectrum of error conditions, making TTS-ASR synthesis a foundation for developing advanced correction systems that address both localized and generalized ASR inaccuracies.

Numerous studies have leveraged TTS–ASR synthesis to generate paired noisy–clean text for specialized ASR refinement tasks, often utilizing multi-speaker TTS systems and diverse augmentation techniques. 
For instance, Wang et al.~\cite{Wang2022Mar} employed a multi-speaker, multi-locale TTS system to synthesize contextual spoken phrases (e.g., names) and their corresponding ASR hypotheses, paired these with the original text, and embedded the resulting pairs into sentence patterns to train a contextual spelling-correction model.
Similarly, Tang et al.~\cite{tang2025full} developed the ChFT dataset for full-text Chinese error correction using a TTS–ASR pipeline built on ChatTTS and a cascaded ASR system, incorporating voice activity detection and punctuation restoration to cover multiple domains.
Gu et al.~\cite{Gu2024May} further extended this approach by combining multi-speaker TTS (e.g., YourTTS) with spectrogram-level augmentations such as frequency masking to produce large-scale synthetic corpora for training a denoising LM, achieving state-of-the-art performance on LibriSpeech. In a related effort, Wang et al.~\cite{Wang2021Aug} focused on context-specific scenarios by synthesizing speech containing placeholders (e.g., names) and performing data augmentation through word replacement to address data sparsity.
Collectively, these studies demonstrate how TTS–ASR synthesis, empowered by multi-speaker systems and innovative augmentation schemes, plays a pivotal role in advancing ASR error correction across diverse domains and applications.

\subsection{Pseudo Data Generation}
Given the scarcity of large-scale, ready-to-use paired datasets for ASR refinement, pseudo-data generation, owing to its largely automated construction process, has become a cornerstone for training robust refinement models. Previous studies~\cite{leng2021fastcorrect, du2022cross, leng2021fastcorrect2} have demonstrated the effectiveness of synthetic data in enhancing ASR refinement performance.
Early efforts, such as Wang et al.~\cite{Wang2020EntityRetrieval}, constructed benchmark datasets from public text corpora, including the Amazon Review~\cite{He2016FebAmazonReviewDataset} and Yelp Open~\cite{YelpDataset} datasets, each containing 100,000 unique entities. They simulated realistic ASR errors using an n-gram confusion matrix derived from commercial dialog systems and ensured generalization by partitioning error patterns between training and testing sets to avoid memorization.

More advanced approaches have leveraged NLMs for pseudo-data creation. Leng et al.~\cite{Leng2022DecSoftCorrect} employed a BERT-based generator to automatically introduce error tokens, incorporating homophone substitutions to better reflect real ASR error types. Their framework trains a 12-layer BERT model with a masked language modeling objective, in which 20\% of tokens undergo noise injection following a fixed distribution: 10\% remain unchanged, 40\% are masked, 40\% replaced with homophones, and 10\% randomly substituted. During generation, the model produces deletion, insertion, and substitution errors by strategically masking tokens and sampling BERT’s predictions as pseudo-errors—yielding more realistic training data than random perturbation methods.
Leng et al.~\cite{leng2021fastcorrect} also generate pseudo-ASR data by injecting realistic substitution, deletion, and insertion errors using homophones and real ASR error distributions.

Hrinchuk et al. \cite{hrinchuk2020correction} further introduced a set of data-augmentation techniques. Their approach injects synthetic noise into ASR hypotheses through operations such as token dropout, token replacement, and token shuffling, encouraging the model to generalize to a wider range of real ASR error patterns. Zhao et al. \cite{zhao2021bart} also suggested three data-augmentation strategies to generate diverse ASR-error patterns for training the semantic correction model: using a less robust acoustic model to produce noisier hypotheses, perturbing acoustic features to introduce variability, and applying n-best expansion by retaining multiple decoding candidates while filtering out overly erroneous ones.

Dutta et al.~\cite{dutta2022error} also created training data in three ways: using an ASR system on out-of-domain data, using a weaker ASR model to induce diverse errors, and synthetically injecting errors by replacing words with similar-sounding ones or those with small edit distance. Wang et al. \cite{wang2022effective} introduce two augmentation techniques: randomly masking alignment tokens at each refinement step and applying SpecAugment \cite{park2019specaugment} to the audio, which together improve the robustness and effectiveness of the Align-Refine decoding process.

Recent work has advanced this line of research by employing LLMs for scalable pseudo-data generation when labeled resources are limited. In EPIC~\cite{Jia2025EPIC}, real ASR error patterns are first extracted from actual outputs and then used to guide LLMs in generating diverse, contextually grounded sentences that mimic realistic ASR mistakes. This LLM-driven strategy facilitates large-scale data augmentation and enhances the robustness of ASR refinement systems, particularly in domain-specific and low-resource scenarios such as classroom dialogue transcription.

\subsection{Data Filtering}
\label{datasets-filtering}
Automatically pairing ASR hypotheses with their reference transcripts is a common approach for constructing ASR refinement datasets. However, such automatically generated data often contains noisy or non-inferable pairs, primarily due to excessive insertion or deletion errors and inaccuracies in the target transcripts. To address this, Udagawa et al.~\cite{udagawa2024robust} introduce a conservative data filtering strategy guided by two principles: (i) a valid correction must enhance linguistic acceptability, and (ii) it must remain inferable from the ASR hypothesis and its phonemic representation. Pairs that violate either condition are retained without modification, allowing the model to avoid unnecessary corrections. This approach effectively reduces overcorrection, particularly in out-of-domain contexts, and outperforms traditional edit-distance-based filtering methods. By prioritizing dataset quality over sheer size, their method demonstrates strong generalization across different LLMs and evaluation domains.

%%%%%%%%%%%%%%%%%%%%%%%%%%%%%%%%%%%%%%%%%%%%%%%%%%%%
%              M E T R I C S
%%%%%%%%%%%%%%%%%%%%%%%%%%%%%%%%%%%%%%%%%%%%%%%%%%%%

\section{Metrics}
\label{sec:metrics}

\begin{figure*}[h]
    
    \centering
    \includegraphics[width=0.75\textwidth, height=0.45\textwidth]{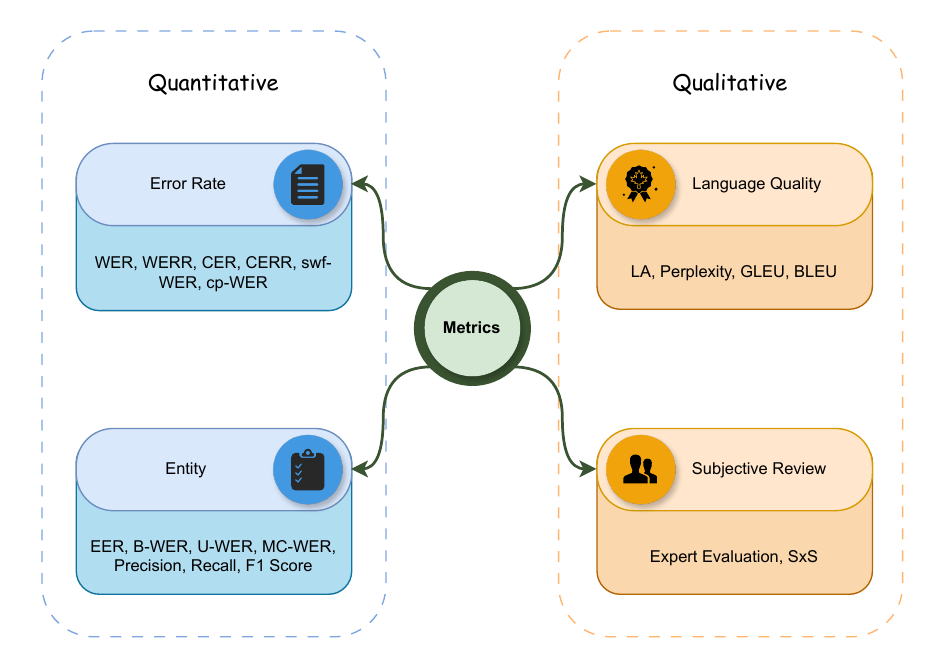}
    \caption{Overview of metrics used to evaluate ASR refinement.}
    \label{fig:metrics}
\end{figure*}

Refinement techniques are essential for enhancing the accuracy and usability of ASR systems and require evaluation metrics that are sufficiently sensitive to key errors. These metrics can be systematically categorized into four classes: error rate metrics, language quality metrics, entity metrics, and qualitative review frameworks. This categorization, detailed below, facilitates the selection of appropriate tools to evaluate ASR refinement methods across various domains of performance, ensuring both precision and applicability in diverse contexts. An overview of this section is provided in Figure \ref{fig:metrics}, and Table \ref{tab:metric_gaps} presents illustrative examples that underscore the necessity of carefully examining evaluation metrics.

\subsection{Error Rate Metrics}
These metrics focus on quantifying transcription accuracy by measuring discrepancies between refined ASR outputs and reference texts. They form the foundation for evaluating transcription correctness at both the word and character levels and are widely used as benchmarks for system improvement. An important consideration is that error-rate metrics typically rely on an alignment process between the hypothesis and reference transcripts. Therefore, the accuracy of this alignment directly affects the validity of the resulting error analysis. While the standard Levenshtein distance is the most commonly used method for alignment, recent studies such as \cite{borgholt2025text, ruiz2015phonetically} have proposed enhanced alignment techniques that better account for phonetic and contextual similarities, offering more reliable assessments of ASR performance. Below, we present a list of widely used error rate metrics.

\begin{itemize}
    \item WER \& WERR: WER measures the percentage of substitutions ($S$), deletions ($D$), and insertions ($I$) required to align the hypothesis and the reference, normalized by the reference length $N$:
    \[
      \mathrm{WER} = \frac{S + D + I}{N}\times100.
    \]
    These error counts are obtained via a standard Levenshtein alignment. 
    WER Reduction (WERR) also quantifies the relative improvement of WER:
    \[
      \mathrm{WERR} = \frac{\mathrm{WER}_{\text{before}} - \mathrm{WER}_{\text{after}}}{\mathrm{WER}_{\text{before}}}\times100.
    \]
    
    \item CER \& CERR: Character Error Rate (CER) applies the same formula at the character level, capturing spelling and morphological mistakes. 
    Character Error Rate Reduction (CERR) is defined analogously to WERR, expressing the percentage reduction in CER.
    \item Stop Word Filtered WER (swf-WER): This variant computes WER after removing common stop words (e.g., ``the,'' ``and'') from both the hypothesis and the reference, 
    focusing the metric on semantically informative terms~\cite{Imai2025Saki}.

    \item Corrected Percentage WER (cpWER): This metric measures the fraction of originally erroneous words that are corrected by a refinement method, 
    highlighting its direct corrective impact~\cite{kirakosyan2024speaker}.
\end{itemize}

\subsection{Entity Metrics}
These metrics target the accuracy of identifying specific entities (e.g., names, dates, or locations) within refined transcriptions, critical for applications requiring structured information extraction.

\begin{itemize}
    \item Entity Error Rate (EER): Quantifies errors in entity recognition, calculated as the proportion of incorrectly identified entities relative to the total.

    \item Biased-WER (B-WER) and Unbiased-WER (U-WER): These are two additional variants of WER that are primarily utilized by adaptation techniques (particularly those detailed in Section \ref{sec:capturing-specific-phrases-words}) to demonstrate the effectiveness of correction methods on a list of biasing entities, while ensuring that performance on non-biasing words is maintained (i.e., does not degrade)~\cite{he2025pmf}. Domain-specific variants also extend the same formulation to domain-specialized unit sets. 
    For example, the Medical Concept WER (MC-WER), a variant of B-WER, introduced in~\cite{adedeji2025multicultural} 
    treats each medical concept as a single atomic word, counting $S_m$, $D_m$, and $I_m$ over $M$ reference concepts:
    \[
      \mathrm{MC\text{-}WER} = \frac{S_m + D_m + I_m}{M}\times100,
    \]
    thereby adapting B-WER to the medical domain without altering its core definition. 
    
    \item Precision: Measures the fraction of proposed corrections that are true positives. In the ASR refinement context, this indicates how often the model’s changes correctly introduce a missing user phrase, rather than hallucinating new content:
    \[
      \mathrm{Precision}
      = \frac{\mathrm{tp}}{\mathrm{tp} + \mathrm{fp}},
    \]
    where $\mathrm{tp}$ is the number of previously missing phrases correctly introduced by the refinement and $\mathrm{fp}$ is the number of phrases erroneously added~\cite{antonova2023spellmapper}.
    
    \item Recall: Measures the fraction of actual missing phrases that the model successfully recovers. This reflects the model’s ability to detect and correct errors in the original ASR output:
    \[
      \mathrm{Recall}
      = \frac{\mathrm{tp}}{\mathrm{tp} + \mathrm{missed}},
    \]
    
    where $\mathrm{missed}$ is the number of originally missing phrases that remain uncorrected and $\mathrm{tp}$ is defined similar to precision metric.
    
    \item F1 Score: The harmonic mean of precision and recall, offering a balanced evaluation of entity recognition performance:
    \[
    \text{F1} = 2 \times \frac{\text{Precision} \times \text{Recall}}{\text{Precision} + \text{Recall}}.
    \]
    
\end{itemize}

\begin{table*}[t]
\centering
\caption{Illustrating gaps and the importance of each metric category for ASR refinement evaluation.}
\label{tab:metric_gaps}
\resizebox{\textwidth}{!}{%
\begingroup
\renewcommand{\arraystretch}{1.35}
\setlength{\tabcolsep}{8pt}
\fontsize{10}{11.5}\selectfont

\begin{tabular}{
p{3.0cm}
>{\centering\arraybackslash}p{3.5cm}
>{\centering\arraybackslash}p{3.5cm}
>{\centering\arraybackslash}p{3.5cm}
>{\centering\arraybackslash}p{3.5cm}
}
\toprule
\textbf{Illustrative Examples} &
\textbf{Example 1} &
\textbf{Example 2} &
\textbf{Example 3} &
\textbf{Example 4} \\
\midrule

\textbf{Ground Truth} &
The cat sat on the mat &
Dr. Smith prescribed antibiotics like amoxicillin for the patient &
The Battle of Hastings was fought in 1066 &
The patient has a history of diabetes \\

\textbf{ASR Transcript} &
A cut start in a math &
Doctor Smith described anti bio things like amoxilin for the patience &
The Battle of Hasting was fought in 1966 &
The patient has a history of diabetics \\

\textbf{Refiner1} &
A cat sat in a mat &
Dr. Smith described antibiotics like amoxicillin for the patient &
The Battle of Hastings was fought in 1966 &
The patient has a history of diabetes \\

\textbf{Refiner2} &
The cut start on the math &
Doctor Smith prescribed anti bio think like amoxilin for the patient &
The Battle of Hasting was in 1066 &
The patient has a history of diabetes and hypertension \\

\textbf{Category with Gap} &
Error Rate Metrics &
Entity Metrics &
Language Quality &
Subjective Review \\

\textbf{Explanation of Gap} &
Both refiners have the same WER (i.e., 3 errors out of 6 words), but Refiner1 corrects more meaningful content words ("cat," "sat," "mat"). WER treats all errors equally and thus misses semantic importance. &
Refiner1 has better entity recognition (correctly identifies "Dr. Smith" and "amoxicillin"), but the verb "described" is contextually wrong. Entity metrics focus on entities and therefore can miss critical verb errors. &
Both refiners produce fluent outputs (similar BLEU scores), but Refiner2 corrects the critical year "1066." Language-quality metrics prioritise fluency and may overlook factual accuracy. &
Refiner2 adds "and hypertension" (a potential hallucination), which may be judged plausible by reviewers without ground truth. Subjective review is subjective and can miss factual errors. \\

\textbf{Complementary Metrics} &
Entity metrics (for key terms like "cat", "mat"); swf-WER (focuses on content words) &
Language-quality metrics (e.g., BLEU for coherence); targeted subjective review (for contextual accuracy) &
Error-rate metrics (e.g., WER detects the substitution); entity metrics (for "1066") &
Error-rate metrics (e.g., WER detects insertions); entity-precision flags hallucinations \\

\bottomrule
\end{tabular}
\endgroup
}
\end{table*}

\subsection{Language Quality Metrics}
These metrics assess the linguistic fluency, naturalness, and coherence of refined ASR transcriptions, moving beyond mere accuracy to evaluate how closely outputs resemble human-like language. They are often employed to gauge usability in real-world applications.

\begin{itemize}
    \item Language Acceptability (LA): It is typically evaluated using an LM, evaluating the fluency and naturalness of transcriptions~\cite{udagawa2024robust}.

    \item Bilingual Evaluation Understudy (BLEU): Originally developed for machine translation, BLEU measures n-gram overlap between ASR output and reference texts, providing a quantitative assessment of fluency and adequacy~\cite{shivakumar2019learning,fathallah2024empowering,li2024crossmodal,mani2020asr}.

    \item GLEU: Designed to address BLEU’s instability on single sentences, GLEU extracts up to 4-grams, computes both precision and recall, and takes their minimum. The resulting symmetric, $[0,1]$-bounded score requires no smoothing and delivers stable per-sentence rewards applicable to reinforcement learning in sequence models~\cite{li2024crossmodal}.

    \item Perplexity: Derived from language modeling, this metric evaluates how well a model predicts a given sample, with lower perplexity indicating better alignment with expected language patterns~\cite{huang2019empirical}.
    
    % \item \textbf{Recall-Oriented Understudy for Gisting Evaluation (ROUGE):} Emphasizes recall by comparing n-gram overlap with reference texts, useful for evaluating information retention in transcriptions.
\end{itemize}

\subsection{Subjective Review Frameworks}
These frameworks address unique evaluation needs, incorporating human judgment to capture aspects not covered by objective categories mentioned earlier.

\begin{itemize}
    \item Expert Evaluation: Involves domain experts reviewing transcriptions for quality, leveraging human insight to assess context-specific nuances~\cite{zhao2021bart}.

    \item Side-by-Side (SxS): A comparative method where human evaluators rate two transcriptions side by side, providing a direct quality comparison~\cite{sainath2019two}.
\end{itemize}

%%%%%%%%%%%%%%%%%%%%%%%%%%%%%%%%%%%%%%%%%%%%%%%%%%%%
%              RESEARCH GAP
%%%%%%%%%%%%%%%%%%%%%%%%%%%%%%%%%%%%%%%%%%%%%%%%%%%%

\section{Research Gap and Future Work}
\label{sec:research-gap}

% text-only methods
Despite substantial advances in ASR refinement, several important gaps remain. One principal concern is the unconstrained use of text-only methods. Many contemporary approaches, particularly those based on general LLMs, assume that ASR output follows the linguistic distributions on which these models were trained. In practice, spoken language (especially informal, conversational speech) exhibits disfluencies, partial words, code-switching, and atypical syntactic patterns. Consequently, applying unconstrained text-only correction can introduce overcorrections or hallucinations that degrade transcription quality, especially when the initial WER is low. Mitigating this risk therefore requires a comprehensive evaluation of conservative, principled correction strategies. Examples include restricting allowable edit types or lengths, grounding edits in phonetic or acoustic evidence, and gating corrections using confidence or uncertainty estimates. Although some methods address parts of these strategies, a comprehensive framework for assessing the individual and combined impact of each strategy, both before and after refinement, remains lacking.

Moreover, while some pipelines incorporate detection modules or confidence estimation to prevent overcorrection, these components are themselves typically trained on text-only data and thus inherit the aforementioned limitations. There is therefore a clear need for more effective mechanisms to guide the correction process.

Importantly, systematic assessment of error types before and after refinement is largely absent: only a small number of studies report post-refinement error-type breakdowns. More research is needed to classify and evaluate errors across domains (e.g., clean vs.\ noisy conditions, general vs.\ domain-specific corpora). %In particular, studying how different modalities — text, phonetics, and raw audio — affect the correction of each error type %
It will help practitioners select the most appropriate refinement strategy given their accuracy requirements and computational constraints.

The use of contextual features has been dominated by text-based signals that can be limited and even misleading especially when WER is high; only a few recent studies exploit contextual information derived directly from speech. Investigating fine-grained acoustic and prosodic features, beyond coarse speech embeddings, represents a promising direction to improve refinement. By leveraging richer speech cues (e.g., phonetic alignments, acoustic tokens), future methods can better ground corrections and reduce erroneous edits.

Another underexplored axis is computational efficiency. Although NAR models and lightweight detection modules have been proposed to reduce latency and cost, relatively few works prioritize or systematically measure implementation complexity and runtime trade-offs. Detailed evaluations that balance effectiveness against inference cost are necessary to ensure that refinement techniques are viable for real-world deployment, where resources and latency budgets are constrained.

A further practical barrier is the difficulty of comparing existing methods without re-implementation: diverse datasets, and evaluation metrics obscure the relative merits of competing approaches. Establishing more unified benchmarking suites or leaderboards that standardize datasets, metrics, and evaluation protocols would enable fair, reproducible comparisons and accelerate progress in the field.

In summary, future research should: (i) develop guided, uncertainty-aware correction strategies that leverage multimodal evidence and assess the usefulness of different modalities (e.g., phonetic features, speech embeddings); (ii) integrate fine-grained speech-context features into the refinement pipeline and evaluate the contribution of each feature type; (iii) provide systematic, error-type–aware evaluations; (iv) prioritize computational efficiency alongside accuracy; and (v) support reproducible benchmarking through standardized evaluations. Addressing these gaps will produce ASR refinement techniques that are both robust and practical for diverse real-world applications.

%%%%%%%%%%%%%%%%%%%%%%%%%%%%%%%%%%%%%%%%%%%%%%%%%%%%
%%%%%%%%%%%%%%CONCLUSION
%%%%%%%%%%%%%%%%%%%%%%%%%%%%%%%%%%%%%%%%%%%%%%%%%%%%

\section{Conclusion}
\label{sec:conclusion}

This survey provides a comprehensive overview of non-intrusive ASR refinement techniques, categorized into five classes: fusion, rescoring, correction, distillation, and training adjustment. It also reviews domain-specific adaptation strategies. By offering a detailed account of dataset creation practices and an extensive taxonomy of evaluation metrics, the survey establishes a practical baseline for future research. The purpose of our structured analysis is generally threefold. First, we aim to help researchers confront challenges in the field by offering a thorough description of the current state and by identifying outstanding research gaps. Second, we seek to guide the selection of appropriate metrics and datasets to enable consistent, comparable evaluations across studies. Third, we intend to motivate the creation of standardized benchmarks and reproducible evaluation protocols,  including leaderboards, to accelerate progress and foster clearer comparisons between existing approaches.

\appendix
\section{Datasets Used in ASR Refinement Literature}
\label{datasets-tables}
Tables \ref{tab:datasets-sr-part1}, \ref{tab:datasets-sr-part2}, and \ref{tab:datasets-sr-part3} provide a list of datasets that were primarily introduced for speech recognition purposes but are widely used in refinement methods.

\begin{table}[h]
\centering
\caption{Datasets Primarily Used for Speech Recognition and Applicable in Refinement Techniques (Part~1). Language abbreviation in the table is: EN (English).}
\label{tab:datasets-sr-part1}
\resizebox{\textwidth}{!}{%
\begingroup
\renewcommand{\arraystretch}{1.3}
\setlength{\tabcolsep}{6pt}
\fontsize{8}{9}\selectfont

\begin{tabular}{l c p{1.6cm} c p{1.8cm} p{4.2cm} c}
\toprule
\textbf{Dataset} &
\textbf{Year} &
\textbf{Lang.} &
\textbf{Primary Task} &
\textbf{Duration} &
\textbf{Notes} &
\textbf{Ref.} \\
\midrule

WSJ & 1993 & EN & ASR & 80 hours &
Read speech from WSJ, news domain &
\cite{garofolo1993csr} \\

ATIS & 1993 & EN & ASR & 34 hours &
Spoken language, travel domain &
\cite{Hemphill1990ATIS} \\

LVCSR & 1996 & EN & ASR & 40 calls &
Conversational speech, telephone calls &
\cite{young1996review} \\

Switchboard & 1997 & EN & ASR & 260 hours &
Conversational speech, telephone calls &
\cite{godfrey1997switchboard} \\

CallHome & 1997 & EN & ASR & 18.3 hours &
Conversational telephone speech, only 5--10 minute segments are transcribed per call &
\cite{canavan1997callhome} \\

Fisher & 2004--2005 & EN & ASR & 1,960 hours &
Conversational speech, telephone calls, manual transcript &
\cite{cieri2004fisher} \\

AMI & 2005 & EN & ASR & 100 hours &
Meetings, multi-speaker, manual transcript &
\cite{kraaij2005ami} \\

CHiME-1 & 2011 & EN & ASR & 19.7 hours &
Read speech (plus noise and RIR) from WSJ, news, manual transcript &
\cite{barker2011chime1} \\

TED-LIUM 1 & 2012 & EN & ASR & 118 hours &
Conference speech, auto alignment &
\cite{rousseau2012ted} \\

CHiME-2 & 2013 & EN & ASR & 324 hours &
News, manual transcript &
\cite{vincent2013second} \\

TED-LIUM 2 & 2014 & EN & ASR & 207 hours &
Conference speech, manual and auto alignment transcript &
\cite{rousseau2014enhancing} \\

LibriSpeech & 2015 & EN & ASR & 1,000 hours &
Read speech from audiobooks &
\cite{7178964} \\

CHiME-3 & 2015 & EN & ASR & 13.3 hours &
Noisy locations (e.g., bus, street), news, manual transcript &
\cite{barker2015third} \\

CHiME-4 & 2016 & EN & ASR & 13.3 hours &
News, same as CHiME-3 but multi-channel, manual transcript &
\cite{Vincent2016Chime4} \\

SWC & 2016 & EN & ASR & 395 hours &
Volunteer readers, encyclopedia &
\cite{baumann2019spoken} \\

\bottomrule
\end{tabular}
\endgroup
}
\end{table}

\clearpage

\begin{table}[h]
\centering
\caption{Datasets Primarily Used for Speech Recognition and Applicable in Refinement Techniques (Part~2). Language abbreviations in the table are: EN (English), ML (Multi-Lingual), and ZH (Chinese).}
\label{tab:datasets-sr-part2}
\resizebox{\textwidth}{!}{%
\begingroup
\renewcommand{\arraystretch}{1.3}
\setlength{\tabcolsep}{6pt}
\fontsize{8}{9}\selectfont

\begin{tabular}{l c p{1.6cm} c p{1.8cm} p{4.2cm} c}
\toprule
\textbf{Dataset} &
\textbf{Year} &
\textbf{Lang.} &
\textbf{Primary Task} &
\textbf{Duration} &
\textbf{Notes} &
\textbf{Ref.} \\
\midrule

LRS2 & 2017 & EN & ASR & 224 hours &
Audio-visual speech, news and shows &
\cite{Chung2016NovLRS2} \\

LRS2 & 2018 & EN & ASR & $\sim$224 hours &
Audio-visual speech dataset from BBC television broadcasts; thousands of spoken sentences for lip reading and ASR research &
\cite{afouras2022deep} \\

LRS3 & 2018 & EN & ASR & $\sim$438 hours &
Audio-visual speech dataset from TED and TEDx talks; spoken sentences with aligned transcripts for lip-reading and ASR research &
\cite{afouras2018lrs3} \\

CHiME-5 & 2018 & EN & ASR & 49 hours &
Conversational dinner party, manual transcript &
\cite{barker2018fifth} \\

TED-LIUM 3 & 2018 & EN & ASR & 452 hours &
Conference speech, manual and auto alignment transcript &
\cite{Hlubik2020Tedlium} \\

SWGISpeech & 2021 & EN & ASR & 5,000 hours &
Finance, manual transcript &
\cite{o2021spgispeech} \\

VoxPopuli & 2021 & ML & ASR & 1,820 hours &
Multilingual corpus collected from 2009--2020 European Parliament recordings; 1,791h multilingual and 29h accented English &
\cite{wang2021voxpopuli} \\

GigaSpeech & 2021 & EN & ASR & $\sim$10,000 hours &
Multi-domain English corpus with $\sim$10,000h of high-quality labeled audio and $\sim$40,000h total audio from audiobooks, podcasts, and YouTube; read and spontaneous speech &
\cite{chen2021gigaspeech} \\

PriMock 57 & 2022 & EN & ASR & 8.5 hours &
Primary-care telemedicine domain, manual transcript &
\cite{korfiatis2022primock57} \\

CORAAL & 2023 & EN & ASR & -- &
Corpus of regional African American Language with audio recordings and time-aligned orthographic transcripts from over 220 sociolinguistic interviews; suitable for ASR and sociolinguistic research &
\cite{kendall2023coraal} \\

Libriheavy & 2024 & EN & ASR & 50,000 hours &
Read speech from LibriVox; includes rich transcripts with punctuation, casing, and context; three subsets (small: $\sim$500h, medium: $\sim$5,000h, large: $\sim$50,000h) &
\cite{kang2024libriheavy} \\

AISHELL-1 & 2017 & ZH & ASR & 178 hours &
Read speech, several domains (finance, technology, sports, news) &
\cite{bu2017aishell} \\

AISHELL-2 & 2018 & ZH & ASR & 1,000 hours &
Read speech, several domains &
\cite{du2018aishell} \\

Aidatatang & 2019 & ZH & ASR & 200 hours &
Read speech, mostly mobile records, manual transcript &
\cite{aidatatang200zh} \\

AISHELL-3 & 2020 & ZH & ASR & 85 hours &
General domain, multi-speaker &
\cite{shi2021aishell} \\

AISHELL-4 & 2021 & ZH & ASR & 120 hours &
Conference/meeting speech, multi-channel &
\cite{fu2021aishell} \\

MAGICDATA-READ & 2019 & ZH & ASR & 755 hours &
Scripted read speech from native Mandarin speakers (1080 speakers, indoor mobile recordings) &
\cite{magicdata2019slr68} \\

MagicData-RAMC & 2022 & ZH & ASR & 180 hours &
Spontaneous conversational Mandarin speech (351 multi-turn dialogues, 663 speakers, mobile recordings, rich manual annotation) &
\cite{yang2022open} \\

AISHELL-5 & 2025 & ZH & ASR & 100 hours &
In-car recordings, multi-channel, noise included &
\cite{dai2025aishell} \\

\bottomrule
\end{tabular}
\endgroup
}
\end{table}

\clearpage

\begin{table}[h!]
\centering
\caption{Datasets Primarily Used for Speech Recognition and Applicable in Refinement Techniques (Part~3). Language abbreviations in the table are: JA (Japanese), DE (German), NL (Dutch), and ML (Multi-Lingual).}
\label{tab:datasets-sr-part3}
\resizebox{\textwidth}{!}{%
\begingroup
\renewcommand{\arraystretch}{1.3}
\setlength{\tabcolsep}{6pt}
\fontsize{8}{9}\selectfont

\begin{tabular}{l c p{1.6cm} c p{1.8cm} p{4.2cm} c}
\toprule
\textbf{Dataset} &
\textbf{Year} &
\textbf{Lang.} &
\textbf{Primary Task} &
\textbf{Duration} &
\textbf{Notes} &
\textbf{Ref.} \\
\midrule

CSJ & 2004 & JA & ASR & 658 hours &
Academic presentation speech, Simulated public speaking &
\cite{maekawa2004corpus} \\

ALFFA  & 2005 & Amharic & ASR & 20 hours &
Newspaper/magazine, manual transcript &
\cite{abate2005amharic} \\

ALFFA  & 2012 & Swahili & ASR & 11.75 hours &
Newspaper/magazine, manual transcript &
\cite{gelas2012developments} \\

ALFFA  & 2016 & Wolof & ASR & 16.8 hours &
Newspaper/magazine, manual transcript &
\cite{gauthier2016collecting} \\

SWC  & 2016 & DE & ASR & 386 hours &
Volunteer readers, encyclopedia &
\cite{baumann2019spoken} \\

SWC & 2016 & NL & ASR & 224 hours &
Volunteer readers, encyclopedia &
\cite{baumann2019spoken} \\

% Common Voice & 2020 & 129 languages & ASR & 33,000 hours &
% General domain &
% \cite{ardila2020common} \\

Common Voice & 2020 & ML & ASR & 33,000 hours &
General domain; 129 languages &
\cite{ardila2020common} \\

LibriSpeech & 2020 & ML & ASR & $\sim$50,000 hours &
Multilingual read speech from LibriVox; about 44.5k hours English and about 6k hours across other languages &
\cite{pratap2020mls} \\

SPREDS-U1 & 2023 & ML & ASR & -- &
Multilingual evaluation data for speech recognition; 21 languages from 22 countries/regions with consistent recording conditions and raw transcriptions &
\cite{NICT2023SPREDSU1} \\

FLEURS & 2023 & ML & ASR & 12 hours per language &
Encyclopedia; 102 languages &
\cite{conneau2023fleurs} \\

\bottomrule
\end{tabular}
\endgroup
}
\end{table}

\bibliographystyle{unsrt}
\bibliography{reference}
%% For numbered reference style

\end{document}